\documentclass[preprint,12pt]{elsarticle}

\usepackage{tikz}
\usetikzlibrary{arrows.meta,positioning,calc}
\usepackage{hyperref}
\usepackage{url}
 \usepackage{booktabs}
\usepackage{amsmath,amsthm}
\usepackage{subcaption}
\usepackage{amssymb}
\usepackage{graphicx}
\graphicspath{{figures/}}
\usepackage{algorithm}
\usepackage{algpseudocode}
\usepackage{caption}
\usepackage{amssymb}

\journal{Nuclear Physics B}

\begin{document}

\begin{frontmatter}



\title{Two-Step Diffusion:  Fast Sampling and Reliable Prediction for 3D Keller-Segel and KPP Equations in Fluid Flows}


\author[aff1]{Zhenda Shen}
\ead{zhenda_shen@g.harvard.edu}

\author[aff2]{Zhongjian Wang}
\ead{zhongjian.wang@ntu.edu.sg}

\author[aff3]{Jack Xin}
\ead{jack.xin@uci.edu}

\author[aff4]{Zhiwen Zhang\corref{cor1}}
\ead{zhangzw@hku.hk}
\cortext[cor1]{Corresponding author.}

\affiliation[aff1]{organization={School of Engineering and Applied Science, Harvard University},
            city={Cambridge},
            state={MA},
            postcode={02138},
            country={USA}}

\affiliation[aff2]{organization={Division of Mathematical Sciences, Nanyang Technological University},
            addressline={21 Nanyang Link},
            postcode={637371},
            country={Singapore}}

\affiliation[aff3]{organization={Department of Mathematics, University of California, Irvine},
            city={Irvine},
            state={CA},
            postcode={92697},
            country={USA}}

\affiliation[aff4]{organization={Department of Mathematics, The University of Hong Kong},
            addressline={Pokfulam Road},
            city={Hong Kong},
            country={China}}

\begin{abstract}
We study fast and reliable generative transport for the 3D KS (Keller-Segel) and KPP (Kolmogorov-Petrovsky-Piskunov) equations in the presence of fluid flows with the goal to approximate the map between initial and terminal distributions for a range of physical parameters $\sigma$ under the Wasserstein metric. 
To minimize the inaccuracy of \emph{direct} Wasserstein solver, we propose a two-stage pipeline that retains one-step efficiency while reinstating an explicit $W_2$ objective where it is tractable. In \textbf{Stage I}, a Meanflow-style regressor 
yields a deterministic, one-step global transport that moves particles close to their terminal states. 
In \textbf{Stage II}, we freeze this initializer and train a near-identity corrector (\emph{Deep Particle, DP}) that \emph{directly} minimizes a mini-batch $W_2$ objective using warm-started optimal transport couplings computed on the Meanflow outputs. Crucially, after the one-step transport (from Stage I) concentrating mass on the approximated correct support, the induced geometry stabilizes high-dimensional $W_2$ computation of the direct Wasserstein solver. We validate our construction in the 3D KS and KPP equations subject to fluid flows with ordered and chaotic streamlines. 
\end{abstract}


\begin{highlights}
\item A two-stage solver for Keller-Segel (KS) and KPP Equations: a 1-NFE (number of function evaluation) Meanflow initializer for a global transport followed by a near-identity DP (DeepParticle) corrector that directly minimizes a mini-batch 2-Wasserstein (\(W_2\)) objective with warm-started optimal transport (OT) couplings.
\item A demonstration that such decomposition turns an intractable high-dimensional \(W_2\) problem into a tractable one: the first step simplifies geometry; the second step performs local, measure-aware alignment without heavy global OT.
\item Empirical gains across a wide range of physical parameters, especially in the singular perturbation regime of KS and KPP, achieving reduction of \(W_2\) values while maintaining 
low NFE and robustness.
\end{highlights}


\end{frontmatter}


\section{Introduction}
\noindent Learning parameterized particle transports that faithfully match simulator outputs while
remaining fast at inference is a core challenge at the intersection of flow-based generative
modeling~\cite{lipman2023flow,liuflow,albergo2023building} and scientific machine learning.
This task becomes especially daunting when the underlying particle system exhibits
chaotic behavior in high-dimensional spaces. The Keller-Segel systems were introduced ~\cite{keller1970initiation} in the 1970s to describe the aggregation of the slime mold Dictyostelium discoideum in response to an attractive chemical signal, yet became harder to simulate 
in the presence of complex fluid flows. Similarly, the Kolmogorov--Petrovsky--Piskunov (KPP) equation describes the evolution of a reaction--diffusion process, exhibiting multi-scale traveling wave behavior in fluid flows~\cite{xin2009introduction}.  In the setting treated here, our goal is to map the initial particles \(x_0\) to terminal states \(x_1\) across a range of physical parameters \(\sigma\) in the distribution sense, without evaluating the physical particle interactions in the intermediate times. The quadratic Wasserstein distance (\(W_2\)) is an attractive choice for this purpose~\cite{peyre2019computational}, yet \emph{directly} optimizing \(W_2\) in high dimensions is notoriously difficult~\cite{genevay2016stochastic}: computing couplings is expensive, gradients are noisy~\cite{genevay2019sample}, and naive mini-batch surrogates can be unstable and biased~\cite{nguyen2022improving}.\\
Recently, few-/one-step flow-based methods make impressive progress on fast, deterministic sampling by regressing velocity fields from paired endpoints without simulating forward diffusions~\cite{salimans2022progressive,geng2023one}. These approaches scale well to high-dimensional data because their learning objective no longer directly minimizes the discrepancy metrics (for instance, Wasserstein distance ~\cite{DP_2022,
wang2024deepparticle,arjovsky2017wasserstein}, KL~\cite{papamakarios2021normalizing}, JS~\cite{goodfellow2020generative} etc) between the generated and the target distributions. A recent state-of-the-art method is the Meanflow (~\cite{geng2025mean}, 2025) method which distills the reverse-time diffusion trajectory into a single flow-matching step by regressing a velocity field from paired endpoints. As a result of the indirect connection between flow matching and target distribution discrepancy, although its one-step predictions often land near the correct support~\cite{baptista2025memorization,lu2023mathematical}, residual misalignment in mass placement and anisotropy may persist, particularly in stiff, multi-scale regimes.\\
To tackle this, we propose a two-stage pipeline that preserves the efficiency and stability of one-step flows while restoring an explicit \(W_2\) training signal where it matters most. In \textbf{Stage I} (Algorithm~\ref{alg:meanflow_train}), we train a Meanflow-style regressor using the Meanflow identity to obtain a \emph{deterministic, 1-NFE} (number of function evaluations) global transport that moves particles close to terminal states across physical parameters. In \textbf{Stage II} (Algorithm~\ref{alg:dp_refine}), we fix the Meanflow generator as the prior generator and train a near-identity corrector (\emph{Deep Particle, DP}) whose loss \emph{directly minimizes a mini-batch approximation to \(W_2\) distance}. The key observation is that after the fast one-step transport In Stage I  has concentrated mass on (or near) the correct support, the induced geometry makes \(W_2\) supervision easy to optimize: couplings are closer to permutation, costs are locally well-conditioned, mini-batch OT becomes stable, and batched
operations with small tensors are GPU-friendly. Thus, instead of attacking a difficult, high-dimensional \(W_2\) optimization from scratch, we first simplify the geometry with a one-step flow, then solve a much easier, local \(W_2\) refinement. This two-step design \emph{avoids a heavy and global OT computation} while giving us a principled, geometry-aware objective.
\medskip


\section{Related Work}~\label{sec:related}
\paragraph{Diffusion Model and Flow Matching Methods}
Classical \emph{diffusion models}~\cite{sohl2015deep, song2019generative, ho2020denoising} learn a score field $\nabla_z \log p_t(z)$ by denoising score matching while simulating (or reversing) a noising SDE~\cite{ho2020denoising}; samples are generated by integrating a reverse \emph{stochastic} process, typically requiring quite a few  function evaluations (NFEs), see \cite{SharpLipDM_2025} for a recent well-posedness and error analysis of generation. Later work (e.g., consistency~\cite{luo2023diff,song2023consistency}/ODE distillation~\cite{zhou2024score,yin2024one}) compresses sampling to a few steps but often relies on auxiliary training stages. In contrast, Flow Matching (FM)~\cite{lipman2023flow,albergo2023building} learns a deterministic velocity field $v_\theta(z,t)$ via supervised regression on analytically known conditional velocities along simple paths between data $x_0$ and prior $x_1$ (e.g., a linear interpolation). The key identity is that the unknown marginal velocity equals the conditional expectation of conditional velocities, enabling stable one-stage training and sampling without ever estimating scores or simulating an SDE~\cite{lipman2023flow}.\\
In the standard FM formulation, a data sample $x_0$ is drawn from $\pi_0$ and a prior sample $x_1$ is drawn from $\pi_1$ (typically $\pi_1=\mathcal{N}(0,I)$). A time-indexed state $z_t$ for $t\in[0,1]$ is then evolved deterministically: as time increases, the state moves according to the velocity given by the learned field $v_\theta(z,t)$ evaluated at the current location and time. For supervision, FM specifies a simple conditional trajectory connecting the two endpoints: at each time $t$, the conditional point $z_t$ is defined by linearly interpolating between $x_0$ and $x_1$, so that the path starts at $x_0$ at $t=0$ and reaches $x_1$ at $t=1$.\\
Since the conditional path is linear in time, its instantaneous conditional velocity is the time derivative of the interpolation, which is constant along the path and equals the endpoint difference $x_1-x_0$. The corresponding \emph{marginal} velocity at a location $z$ and time $t$ is obtained by averaging these conditional velocities over all pairs $(x_0,x_1)$ whose interpolated point at time $t$ equals $z$; equivalently, it is the conditional expectation of the conditional velocity given the event $z_t=z$. This identity implies that regressing $v_\theta(z,t)$ onto conditional targets induced by simple paths is sufficient to recover the correct marginal velocity field, enabling one-stage training and deterministic sampling.

\paragraph{One-Step Flow-Matching Model via Meanflow}
 Standard Flow Matching models~\cite{lipman2023flow} learn the marginal velocity field $v^*(z,t)$, which typically induces \emph{curved} Eulerian trajectories; when combined with a \emph{low-NFE} numerical solver, the ODE integration can be \emph{inaccurate}, leading to endpoint bias and distributional mismatch (often exacerbated by guidance). By contrast, Meanflow~\cite{geng2025mean} targets the \emph{average (endpoint) displacement} directly and leverages an analytical mean--flow identity for supervision, yielding a natural one-step sampler that avoids time integration and is less sensitive to trajectory curvature.\\
For $0\le r<t\le 1$, the average velocity is defined as:
\begin{equation}
  u(z_t,r,t)
  =
  \frac{1}{\,t-r\,}\int_{r}^{t} v(z_\tau,\tau)\,d\tau .
  \label{A1}
\end{equation}
As $r\rightarrow t$, $u(z_t,r,t)\to v(z_t,t)$. Moreover, additivity of the integral implies the \emph{consistency relation~\cite{song2023consistency}}: for any $r<s<t$,
\begin{equation}
 (t-r)u(z_t,r,t)
 =
 (s-r)u(z_s,r,s) + (t-s)u(z_t,s,t).
 \label{A2}
\end{equation}
Rewriting \eqref{A1} as $(t-r)u(z_t,r,t)=\int_r^t v(z_\tau,\tau)\,d\tau$ and differentiating
w.r.t.\ $t$ (with $r$ fixed) gives
\begin{equation}
  u(z_t,r,t)
  =
  v(z_t,t) - (t-r)\frac{d}{dt}u(z_t,r,t),
  \label{A3}
\end{equation}
Equations \eqref{A1} and \eqref{A3} are equivalent, and the equation \eqref{A3} is the \emph{Mean--Flow Identity~\cite{geng2025mean}}.\\
The Meanflow method offers 1-NFE and deterministic sampling with simple conditioning and well-conditioned averaged-velocity targets, but can retain residual bias on multi-modal/multi-scale data, is sensitive to time sampling and JVP accuracy, and often benefits from a lightweight corrector (e.g., OT refinement~\cite{tong2023improving}).

\paragraph{Wasserstein distance and its discretizations}

The Wasserstein distance provides a geometry on probability measures by quantifying the minimal ``effort'' required to morph one distribution into another~\cite{villani2021topics}. For probability measures \(\mu\) and \(\nu\) on a metric space \((Y,\mathrm{dist})\) and \(p=2\), the quadratic Wasserstein distance is

\begin{equation}
W_2(\mu,\nu)
=\Bigg(\inf_{\gamma\in \Gamma(\mu,\nu)}
\int_{Y\times Y}\mathrm{dist}(y',y)^2\,\mathrm{d}\gamma(y',y)\Bigg)^{1/2},
\end{equation}
where \(\Gamma(\mu,\nu)\) is the set of couplings (joint measures) with marginals \(\mu\) and \(\nu\). When a map \(f:X\to Y\) pushes \(\mu\) forward to \(f_\ast\mu\), the distance between the transformed source and target can be written as
\begin{equation}
W_2\big(f_\ast\mu,\nu\big)
=\Bigg(\inf_{\gamma\in \Gamma(\mu,\nu)}
\int_{X\times Y}\mathrm{dist}(f(x),y)^2\,\mathrm{d}\gamma(x,y)\Bigg)^{1/2}.
\end{equation}
For computation~\cite{peyre2019computational}, one typically replaces \(\mu\) and \(\nu\) by empirical measures with \(N\) samples,
$
\mu=\tfrac1N\sum_{i=1}^N\delta_{x_i},
\nu=\tfrac1N\sum_{j=1}^N\delta_{y_j}.
$
In this discrete setting, a coupling \(\gamma\) becomes a nonnegative \(N\times N\) matrix \((\gamma_{ij})\) with unit row/column sums (a doubly stochastic matrix~\cite{sinkhorn1964relationship}),

\begin{equation}
\gamma_{ij}\ge 0,\quad
\sum_{i=1}^N \gamma_{ij}=1,\quad
\sum_{j=1}^N \gamma_{ij}=1,
\end{equation}
so the discrete quadratic Wasserstein objective reduces to the linear program

\begin{equation}
\widehat{W}_2(f)
=\Bigg(\inf_{\gamma\in\Gamma^N}
\frac{1}{N}\sum_{i=1}^N\sum_{j=1}^N
\mathrm{dist}\big(f(x_i),y_j\big)^2\,\gamma_{ij}\Bigg)^{1/2}.
\end{equation}
Here, \(\gamma_{ij}\) represents the proportion of mass moved from \(f(x_i)\) to \(y_j\); the optimal value is the minimum transport ``effort'' needed to align \(f_\ast\mu\) with \(\nu\). In practice, this discrete formulation underlies modern optimal transport solvers and learning pipelines, often estimated on mini-batches and accelerated via entropic regularization and Sinkhorn iterations~\cite{altschuler2017near}, which trade exactness for speed while preserving the geometric bias of optimal transport.

\section{Methodology}
\noindent We now describe our two–stage pipeline for fast and accurate transport between particle
distributions. \textbf{Stage~I} uses a Meanflow–style model to learn a deterministic, one–step
transport map from the prior to the terminal distribution in Section~\ref{subsec:stage1}. \textbf{Stage~II} in Section~\ref{subsec:stage2} then refines these outputs using a near–identity Deep Particle (DP) corrector trained with a mini–batch
approximation of the quadratic Wasserstein distance \(W_2\). A schematic overview is shown
in Figure~\ref{fig:flow_chart}.

\subsection{Stage I: Meanflow as a one–step generative process}
\label{subsec:stage1}
\noindent Meanflow~\cite{geng2025mean}, in the Flow–Matching view, is modeling a velocity field $v(z,t)$ which transports the prior at $t{=}1$ to the data at $t{=}0$.
Along the interpolation $z_t=(1-t)\,x_0+t\,x_1$, The endpoint relation is
\begin{equation}
  x_{0} = x_{1} - \int_{0}^{1} v(z_{t},t)\,dt. \label{eq:mf-int}
\end{equation}
Define the \emph{average (mean) velocity} between $(r,t)=(0,1)$, as in Equation~\ref{A1} by
\begin{equation}
  u(x_{1},0,1) = \int_{0}^{1} v(z_{t},t)\,dt. \label{eq:mf-avg01}
\end{equation}
Combining \eqref{eq:mf-int}–\eqref{eq:mf-avg01} yields the one–step sampling map
\begin{equation}
  x_{0} = x_{1} - u(x_{1},0,1) = T(x_{1}). \label{eq:mf-onestep-map}
\end{equation}
In practice, we train a network $u_\theta$ to approximate $u(\cdot,0,1)$ and use the single forward pass
$\hat x_0=\varepsilon-u_\theta(\varepsilon,0,1,\sigma)$ for sampling, as shown in Algorithm~\ref{alg:meanflow_train}.\\
Although the Meanflow is less sensitive to trajectory curvature~\cite{geng2025mean} than direct few-step ODE integration (it predicts the path integral in \eqref{eq:mf-avg01} directly), it remains within the Flow–Matching paradigm and exhibits several sources of error when used alone:
(i) {Curvature bias:} the marginal field $v(\cdot,t)$ typically induces \emph{curved} characteristics; learning a single displacement $u(\cdot,0,1)$ tends to capture coarse global transport but smooths out fine structures.
(ii) {Path mismatch:} the linear training path $z_t$ need not coincide with the true characteristics of the marginal flow; the learned average may therefore underfit localized details.
These effects are amplified in AI for Science settings, where data are low-dimensional but high accuracy is required; recovering the distributional detail may demand disproportionately long training or larger models.
\subsection{Stage II: Wasserstein refinement (bringing distributions closer, then aligning)}
\label{subsec:stage2}
\noindent To correct the residual bias from the one–step map \eqref{eq:mf-onestep-map}, we refine the MF outputs by \emph{explicitly} minimizing a Wasserstein distance between the generated and reference distributions. $W_2$ provides a geometry–aware metric that measures distributional misalignment even when supports do not overlap, making it well-suited for precise alignment in low dimensions.
However, its computation is expensive:
(i) constructing the pairwise cost matrix is $\mathcal{O}(N^2)$ in both time and memory for a batch of size $N$;
(ii) Exact solvers scale superlinearly, and entropic or accelerated variants (e.g., Sinkhorn) require many iterations when the two distributions are \emph{far apart} or when small regularization is needed for high accuracy.
Thus, a naive end–to–end $W_2$ minimization from a poor initialization can be prohibitively slow.\\
The refinement map is approximated as a \emph{near–identity} ResNet~\cite{he2016deep}, which is well conditioned for small corrections and empirically attains higher accuracy than unrestricted parameterization. To improve the computation efficiency, we employed the deep particle method ~\cite{wang2024deepparticle}, which uses mini-batch OT and a randomized block interior point solver\cite{xie2024randomized} to approximate the Wasserstein distance. Given mini-batches $\{x_i^{\mathrm{mf}}\}$ and $\{x_j^{\mathrm{ref}}\}$, we form $C_{ij}=\|f_\phi(x_i^{\mathrm{mf}},\sigma)-x_j^{\mathrm{ref}}\|_2^2$ and update a doubly-stochastic coupling $\gamma$ using a standard mini-batch OT routine; the training objective is 
\begin{equation}
    \label{stage2boj}\mathcal{L}_{\mathrm{refine}}=\langle C,\gamma\rangle+\lambda_{\mathrm{res}}\sum_i\|f_\phi(x_i^{\mathrm{mf}},\sigma)-x_i^{\mathrm{mf}}\|_2^2
\end{equation}, which promotes the near-identity structure while aligning residual discrepancies. A similar learning objective promoting OT as the transport map can be found in \cite{li2025dpot}. 
\subsection{Two-Step Design and Training Procedure}
\begin{figure}[htbp]
\label{fig:flowchart}
\centering
\scalebox{1}{
\begin{tikzpicture}[
    >=Latex,
    node distance = 11mm and 18mm,
    box/.style = {draw, rounded corners, minimum width=28mm, minimum height=9mm,
                  align=center, line width=0.6pt},
    every node/.style = {font=\small}
]

\colorlet{mfFill}{blue!7}
\colorlet{mfDraw}{blue!60!black}
\colorlet{pdeFill}{teal!7}
\colorlet{pdeDraw}{teal!60!black}
\colorlet{dpFill}{orange!10}
\colorlet{dpDraw}{orange!75!black}

\colorlet{arrInit}{gray!70!black}
\colorlet{arrMF}{blue!75!black}
\colorlet{arrPDE}{teal!80!black}
\colorlet{arrToDP}{orange!80!black}
\colorlet{arrW2}{red!70!black}
\colorlet{arrFlow}{brown!75!black}
\colorlet{arrOut}{purple!75!black}

\node[box, fill=mfFill, draw=mfDraw] (mf) {Meanflow};
\node[right=12mm of mf, text=arrMF] (xhat0) {$\widehat{x}_0$};
\node[box, fill=dpFill, draw=dpDraw, right=12mm of xhat0] (dp) {Deep particle};
\node[right=12mm of dp, text=arrOut] (xtilde0) {$\widetilde{x}_0$};

\node[box, fill=pdeFill, draw=pdeDraw, below=16mm of mf, xshift=-26mm] (pde) {PDE solver};
\node[right=9mm of pde, text=arrPDE] (x0) {$x_0$};

\node[left=28mm of $(mf)!0.5!(pde)$] (x1) {$x_1$};

\draw[->, line width=0.9pt, arrInit] (x1) |- ($(mf.west)+(-18mm,0)$) -- (mf);

\draw[->, line width=0.9pt, arrInit] (x1) |- (pde);

\draw[->, line width=0.9pt, arrPDE] (pde) -- (x0);
\draw[->, line width=0.9pt, arrMF]  (mf) -- (xhat0);
\draw[->, line width=0.9pt, arrToDP] (xhat0) -- (dp);
\draw[->, line width=0.9pt, arrOut]  (dp) -- (xtilde0);

\draw[->, line width=0.9pt, arrFlow]
  (x0) to[out=70, in=200]
  node[above,pos=0.55, text=arrFlow, fill=white, inner sep=1.2pt, rounded corners=0.8pt]
  {flow-based supervision}
  (mf);

\draw[->, line width=0.9pt, arrW2]
  (x0) to[out=8, in=210]
  node[below,pos=0.55, text=arrW2, fill=white, inner sep=1.2pt, rounded corners=0.8pt]
  {Wasserstein supervision}
  (dp);

\end{tikzpicture}
}
\caption{Two-Step pipeline. Meanflow produces an initial estimator $\widehat{x}_0$ with flow-based supervision, the PDE solver provides $x_0$, and Deep particle refines to $\widetilde{x}_0$ via  Wasserstein-2 Distance supervision.}
\label{fig:flow_chart}
\end{figure}
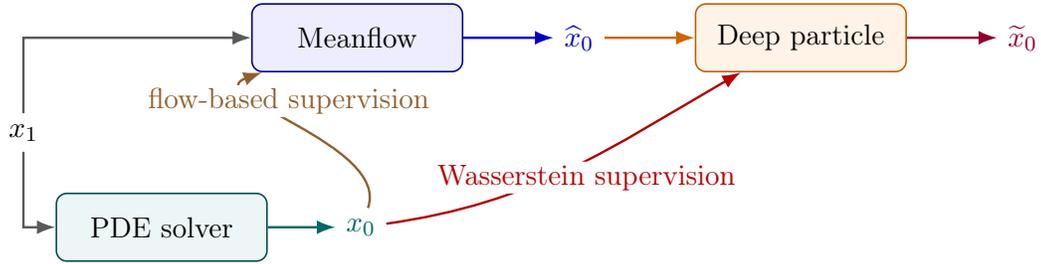

\noindent We therefore adopt a \emph{coarse–to–fine} pipeline which is clearly shown in Figure~\ref{fig:flow_chart}:
\begin{itemize}
  \item \textbf{Stage I (Meanflow)}. Use the one-step generator \eqref{eq:mf-onestep-map} to move samples quickly along the global transport, bringing the model distribution close to the target with {1–NFE} per sample, as in Algorithm~\ref{alg:meanflow_train}.
  \item \textbf{Stage II (Wasserstein refinement).} Starting from these near–aligned particles, minimize a mini–batch approximation of $W_2$ to correct remaining local misalignments. Because the two distributions are now close, the OT coupling converges rapidly with not too many iterations and the per-iteration cost is dominated by a manageable $\mathcal{O}(N^2)$ cost-matrix build. This stage is summarized in Algorithm~\ref{alg:dp_refine}.
\end{itemize}
We train the two modules sequentially. \textbf{Stage I (Meanflow)} regresses the averaged displacement using the meanflow identity. At each iteration we sample paired endpoints $(x_0,x_1,\sigma)$ and times with $r=t$ for half of the batch. We construct the linear path $z_t=(1-t)\,x_0+t\,x_1$ and velocity $v=x_1-x_0$ and then compute a single forward-mode JVP to obtain $(u,\dot u)$. Finally, we form the target $u_{\text{tgt}}=v-(t-r)\,\dot u$ and minimize the regression loss
$
L_{\text{MF}}=\frac{1}{B}\sum\nolimits_{b=1}^B\bigl\|\,u-\mathrm{sg}[\,u_{\text{tgt}}\,]\bigr\|_2^2,
$
where $\mathrm{sg}[\cdot]$ denotes a stop-gradient. After convergence, we freeze $\theta$ and use the one-step map $x_{\text{mf}}=\varepsilon-u_\theta(\varepsilon,0,1,\sigma)$ as a deterministic initializer.
\textbf{Stage II (Wasserstein refinement)} trains a near-identity corrector $f_\phi(x,\sigma)=x+\alpha\,h_\phi(x,\sigma)$ (last layer zero-initialized). For each $\sigma$ we draw fresh mini-batches of $(x_{\text{mf}},x_{\text{ref}})$; every $S_\gamma$ steps we recompute the mini-batch OT coupling $\gamma$ on the current outputs with costs $C_{ij}=\bigl\|f_\phi(x_{\text{mf},i},\sigma)-x_{\text{ref},j}\bigr\|_2^2$. Training alternates parameter updates of $\phi$ with periodic batch refreshes until the validation $W_2$ plateaus. At inference, sampling is 1-NFE and deterministic:
\begin{equation}
    \varepsilon\ \mapsto\ \hat x_0=\varepsilon-u_\theta(\varepsilon,0,1,\sigma)\ \mapsto\ \tilde x_0=f_\phi(\hat x_0,\sigma).
\end{equation}
\begin{algorithm}[h]
\footnotesize
\caption{Stage I: Meanflow Training via Mean--Flow Identity (JVP Regression)}
\label{alg:meanflow_train}
\begin{algorithmic}[1]
\Require  \\
\hspace{1.4em} Paired endpoints $\{(x_0^{(i)},x_1^{(i)},\sigma^{(i)})\}$; \\
\hspace{1.4em} MF network $u_\theta(z,r,t,\sigma)$; \\
\hspace{1.4em} MF batch size $B_{\text{mf}}$, training steps $T_{\text{mf}}$.
\Statex

\State Build endpoint table: $X^{\text{pairs}} = \{(x_0^{(i)},x_1^{(i)},\sigma^{(i)})\}_i$
\For{$k = 1$ to $T_{\text{mf}}$}
  \State Sample mini-batch $(x_0,x_1,\sigma) \sim X^{\text{pairs}}$
  \State Sample times $(t,r)$ with $t \ge r$ (e.g., logit-normal); set $r{=}t$ for half the batch
  \State Define linear path: $z \gets (1{-}t)\,x_0 + t\,x_1$, velocity $v \gets x_1 - x_0$
  \State Compute JVP for total derivative:
  \Statex \hspace{2.6em} $(u,\dot{u}) \gets \mathrm{JVP} \Big((z,t,r)\mapsto u_\theta(z,t,r,\sigma),  (z,t,r); (v,\,1,\,0)\Big)$
  \State Construct Meanflow target via identity $u = v - (t{-}r)\,\frac{d}{dt}u$:
  \Statex \hspace{2.6em} $u_{\text{tgt}} \gets v - (t{-}r)\,\dot{u}$ \hfill (no gradient through $u_{\text{tgt}}$)
  \State Compute loss: $\mathcal{L}_{\text{MF}} = \frac{1}{B_{\text{mf}}}\sum \|u - \mathrm{sg}[u_{\text{tgt}}]\|_2^2$
  \State Update parameters: $\theta \leftarrow \theta - \eta_\theta \nabla_\theta \mathcal{L}_{\text{MF}}$
\EndFor
\State \textbf{Output:} trained Meanflow parameters $\theta$
\end{algorithmic}
\end{algorithm}
\begin{algorithm}[h]
\footnotesize
\caption{Stage II: Wasserstein DP Refinement with Mini-Batch OT}
\label{alg:dp_refine}
\begin{algorithmic}[1]
\Require \\
\hspace{1.4em} Trained Meanflow $u_\theta$; \\
\hspace{1.4em} DP map $f_\phi(x,\sigma) = x + \alpha\,h_\phi([x,\sigma])$ (zero-initialized last layer); \\
\hspace{1.4em} OT solver (e.g., EMD) for mini-batch coupling $\gamma$; \\
\hspace{1.4em} DP batch size $B_{\text{dp}}$, iterations $T_{\text{dp}}$; refresh periods $S_{\text{batch}}, S_\gamma$; near-identity scale $\alpha$.

\Statex
\State \textbf{Step 1: Build DP training pairs using MF outputs}
\For{each conditioning value $\sigma$}
  \State Sample priors $\varepsilon \sim \pi_1(\sigma)$
  \State Generate MF outputs: $x_{\text{mf}} \gets \varepsilon - u_\theta(\varepsilon,r{=}0, t{=}1, \sigma)$
  \State Draw references $x_{\text{ref}} \sim \mathcal{S}(\sigma)$
  \State Store paired mini-batches $\big([x_{\text{mf}},\sigma], x_{\text{ref}}\big)$
\EndFor

\Statex
\State \textbf{Step 2: Train DP corrector via mini-batch OT}
\State Initialize $\phi$ (last layer zeros); $\gamma \gets \tfrac{1}{B_{\text{dp}}}\mathbf{1}\mathbf{1}^\top$ (uniform coupling)
\For{$t = 1$ to $T_{\text{dp}}$}
  \If{$t \bmod S_{\text{batch}} = 0$}
    \State Sample fresh mini-batch $\{x_{\text{mf}},x_{\text{ref}},\sigma\}$ of size $B_{\text{dp}}$
  \EndIf
  \State Compute refined outputs: $x_\phi \gets f_\phi(x_{\text{mf}},\sigma)$
  \If{$t \bmod S_\gamma = 0$}
    \State Compute cost $C_{ij} \gets \|x_{\phi,i} - x_{\text{ref},j}\|_2^2$
    \State Update coupling $\gamma \gets \arg\min_{\gamma\in\Gamma_{B_{\text{dp}}}}\langle C,\gamma\rangle$ 
  \EndIf
  \State Compute OT loss: $\mathcal{L}_{\text{DP}} \gets \langle C,\gamma\rangle +\lambda_{\mathrm{res}}\sum_i\|f_\phi(x_i^{\mathrm{mf}},\sigma)-x_i^{\mathrm{mf}}\|_2^2$
  \State Update parameters: $\phi \leftarrow \phi - \eta_\phi \nabla_\phi \mathcal{L}_{\text{DP}}$
\EndFor
\State \textbf{Output:} trained DP corrector $\phi$
\end{algorithmic}
\end{algorithm}
\noindent This two–stage design avoids (i) \emph{the heavy training burden} Meanflow would incur to recover distributional details on its own, and (ii) \emph{the large computational overhead} of optimizing Wasserstein distance when the two distributions start far apart, while markedly improving final accuracy. The two-stage procedure is summarized in Algorithm~\ref{alg:meanflow_train} and \ref{alg:dp_refine}.\\
This design also offers three main advantages:
(i)\emph{ Fast global transport.} MF offers $1$-NFE sampling and serves as a strong initializer that moves particles close to terminal states, drastically reducing refinement burden.
(ii)\emph{ Stable training signals.} The JVP-based residual with adaptive weighting provides a well-conditioned objective that is robust to outliers and multi-scale dynamics.
(iii)\emph{ Geometry-aware local alignment.} Mini-batch $W_2$ couplings refines MF outputs, where barycentric pulls derived from $\gamma$ correct small residual biases while preserving near-identity structure, largely shortening the gap between the distributions.\\
To be noted, Meanflow cannot be used as a second stage after Deep Particle (DP): Meanflow is restricted to mappings from a Gaussian base to the target, whereas DP transports between arbitrary distributions.
\subsection{Network Architecture}\label{sec:arch}
\noindent Following Stages~I–II, we instantiate two modules consistent with the notation above. The transport predictor $u_\theta:\mathbb{R}^d\times[0,1]^2\times\mathbb{R} \to \mathbb{R}^d$ takes $z$ concatenated with sinusoidal embeddings of $(r,t,\sigma)$, where ${\sigma}$ is the embedded physical parameter, and is implemented as a compact MLP (width 64, depth 5, SELU, light skip connections) with a small-gain linear head; during training we evaluate $(u,\dot u)$ via a single forward-mode JVP along the direction $(v,1,0)$ to obtain $\dot u=\partial_t u(z,r,t)+(\nabla_z u(z,r,t))\,v$, as required by \eqref{A3}, while sampling uses the endpoint map in \eqref{eq:mf-onestep-map} with $u_\theta(\cdot,0,1,\sigma)$. The refinement map $f_\phi:\mathbb{R}^d\times\mathbb{R} \to \mathbb{R}^d$ is parameterized as a near-identity ResNet, $f_\phi(x,\sigma)=x+\alpha\,h_\phi([x,\sigma])$, depth 4, width 64, SiLU activations and LayerNorm, to focus capacity on small corrections introduced in \textbf{Stage~II}. 

\section{Experiments}\label{sec:experiments}
\noindent This section evaluates the two-step pipeline under increasing flow complexity and dimension. Section~\ref{subsec:kslaminar} studies the 3D Keller--Segel system in a steady laminar shear and sweeps the advection amplitude $\sigma$, reporting empirical $W_2$ and qualitative slice projections. Section~\ref{subsec:ksKol} repeats the study under a Kolmogorov flow with chaotic streamlines to stress anisotropy and out-of-range generalization. Section~\ref{subsec:kpp} turns to KPP front-speed estimation on the two-torus via a Feynman--Kac particle system, comparing cold starts against MF and MF+DP warm starts using the eigenvalue estimator $\hat c_T(\lambda)=\hat\mu_T(\lambda)/\lambda$ and invariant-measure $W_2$. Section~\ref{subsec:3dkpp} extends to the fully 3D, time-dependent Kolmogorov flow, assessing the same metrics via 2D projections of the 3D empirical measure. Across subsections, we train on a subset of physical parameter $\sigma$ values and report both interpolation and extrapolation performance.

\subsection{3D Keller-Segel System in Laminar Flow}
\label{subsec:kslaminar}
\noindent In this experiment, we simulate particle density evolution under chemotaxis and a steady 3D laminar flow.  In reduced form, the bacteria density function $\rho$ in the elliptic-parabolic Keller-Segel system evolves according to (\cite{wang2024deepparticle} and references therein):
\begin{equation}
    \rho_t +\textbf{v}\cdot \nabla \rho  = \mu\,\Delta \rho
         + \chi\,\nabla\cdot\bigl(\rho\,\nabla(K*\rho)\bigr)
         \label{ks}
\end{equation}
where $\mu, \chi$ are non-negative constants, and $K(\textbf{x}) = \tfrac{1}{4\pi}\|\textbf{x}\|^{-1}$ is the Green’s function of the 3D Laplacian with $\textbf{x}=(x,y,z)$.  
The background velocity is chosen as a divergence–free laminar (shear) flow:
\begin{equation}
    \textbf{v}(x,y,z) =\sigma\,\bigl(e^{-y^2 - z^2},\,0,\,0\bigr)^T,
\end{equation}
where $\sigma$ is the flow strength along $x$-axis. The ground–truth particle dynamics are generated by an interacting particle approximation of \eqref{ks} in the large particle number limit, i.e.,
\begin{equation}
    dX_j = -\frac{\chi}{J}\sum_{i\neq j}\nabla K_\delta(|X_i-X_j|)\,dt
         + \textbf{v}(X_j)\,dt
         + \sqrt{2\mu}\,dW_j,\;\; J\gg 1,
\end{equation}
where $K_\delta(z): =K(z)\frac{|z|^2}{|z|^2+\delta^2}$ is a regularized kernel with a small parameter $\delta >0 $.\\
We generate training pairs $(x_0, x_1)$ under the 3D laminar flow setting by simulating particle ensembles under different advection amplitudes $\sigma$.  
Specifically, we take $n_{\text{dict}}=8$ logarithmically spaced amplitudes $\sigma$ in the range $[15,150]$, defined by
\[
\sigma_k = 1.5 \times 10^{\,1+\frac{k-1}{\,n_{\text{dict}}-1\,}}
= 15\,\big(10^{1/(n_{\text{dict}}-1)}\big)^{k-1},\qquad k=1,\dots,n_{\text{dict}}.
\]
and collect $N_{\text{particles}}=15{,}000$ samples in $\mathbb{R}^3$ for each choice of $\sigma$.  
The forward simulations runs up to $t=0.02$ with step size $\Delta t = 5{\times}10^{-3}$.  
The initial data $x_0$ is shared across all $\sigma$, and the terminal state at $t=0.02$ forms the target data $x_1$.\\
\begin{figure}[h]
\centering
\begin{minipage}{0.35\linewidth}
\centering
\scriptsize
\begin{tabular}{lcc}
\toprule
\textbf{$\sigma$} & \textbf{Meanflow} & \textbf{DP refinement} \\
\midrule
$20^{(\blacktriangle)}$  & 0.0084 & \textbf{0.0047} \\
$40^{(\blacktriangle)}$  & 0.0068 & \textbf{0.0046} \\
$60^{(\blacktriangle)}$  & 0.0070 & \textbf{0.0049} \\
$80^{(\blacktriangle)}$  & 0.0105 & \textbf{0.0056} \\
$100^{(\blacktriangle)}$ & 0.0132 & \textbf{0.0059} \\
$120^{(\blacktriangle)}$ & 0.0201 & \textbf{0.0065} \\
$140^{(\blacktriangle)}$ & 0.0216 & \textbf{0.0073} \\
$160^{(\bullet)}$        & 0.0403 & \textbf{0.0082} \\
$180^{(\bullet)}$        & 0.0832 & \textbf{0.0140} \\
$200^{(\bullet)}$        & 0.1970 & \textbf{0.0214} \\
\bottomrule
\end{tabular}
\end{minipage}
\hfill
\begin{minipage}{0.54\linewidth}
\centering
\includegraphics[width=0.95\linewidth]{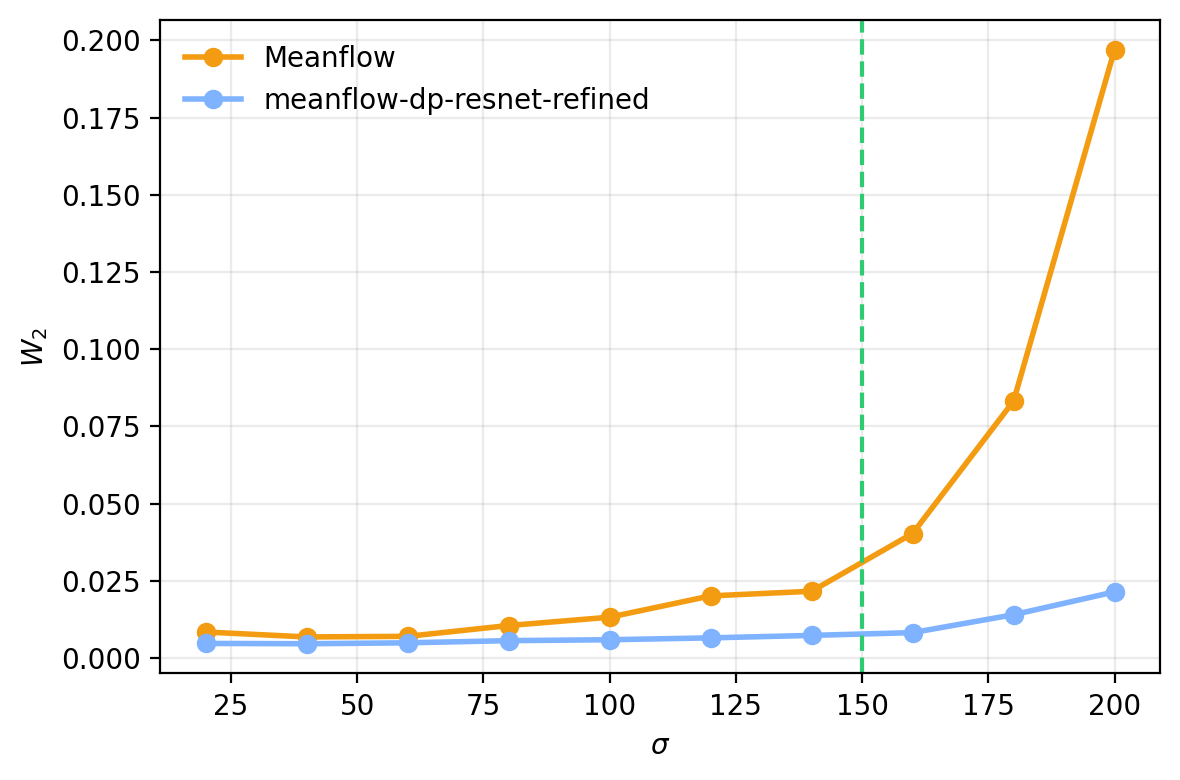}
\end{minipage}

\caption{%
$W_2$ across $\sigma$ (left table) and $W_2$ vs.\ $\sigma$ (right). 
Superscripts on $\sigma$ denote usage: 
$\blacktriangle$—\emph{interpolation}, 
$\bullet$—\emph{extrapolation}. 
DP refinement consistently achieves the lowest cost and shows the largest gains in the singular perturbation regime ($\sigma\!\gtrsim\!150$). 
MF uses one-shot sampling; DP is the refinement in Algorithm~\ref{alg:dp_refine}.}
\label{fig:w2_combo}
\end{figure}
\\Across the full sweep of $\sigma$, MF+DP consistently achieves a lower quadratic Wasserstein distance $W_2$ than Meanflow alone. The reduction in the interpolation regime is modest but systematic. In contrast, once $\sigma$ moves beyond the training range, Meanflow’s error increases rapidly, while MF+DP remains stable. This trend is evident in Fig.~\ref{fig:w2_combo}, where the Meanflow curve rises sharply for $\sigma > 150$ but the refined model stays nearly flat. These results support a two-stage strategy in which a one-step Meanflow transport provides global displacement and a lightweight optimal-transport corrector removes the remaining distributional mismatch. At $\sigma = 160$, an extrapolation point, the reference solution exhibits anisotropic advection along $x$: mass near $(y,z)\approx(0,0)$ is transported farther in $x$ than mass at larger $|y|$ or $|z|$. This produces a tapered pattern in the x-y and x-z projections and an approximately isotropic footprint in y-z. One-step Meanflow captures the overall displacement but remains overly diffuse and biased toward $x$. After applying the DP corrector, the distribution contracts toward the reference barycenter and the anisotropic taper is recovered, with sharper central gradients and a more circular y-z projection, without visible artifacts. The agreement at $\sigma = 160$ indicates robust out-of-distribution generalization of the proposed two-stage pipeline.

\begin{figure}[H]
  \centering
  \begin{subfigure}{0.95\linewidth}
    \centering
    \includegraphics[width=\linewidth]{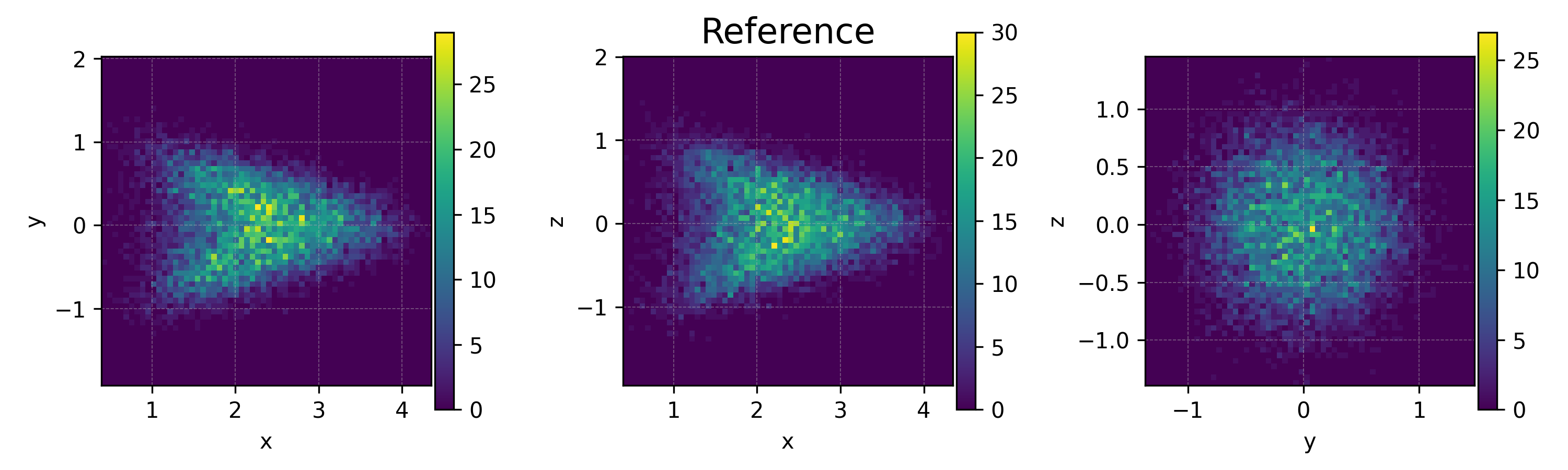}
  \end{subfigure}
  \begin{subfigure}{0.95\linewidth}
    \centering
    \includegraphics[width=\linewidth]{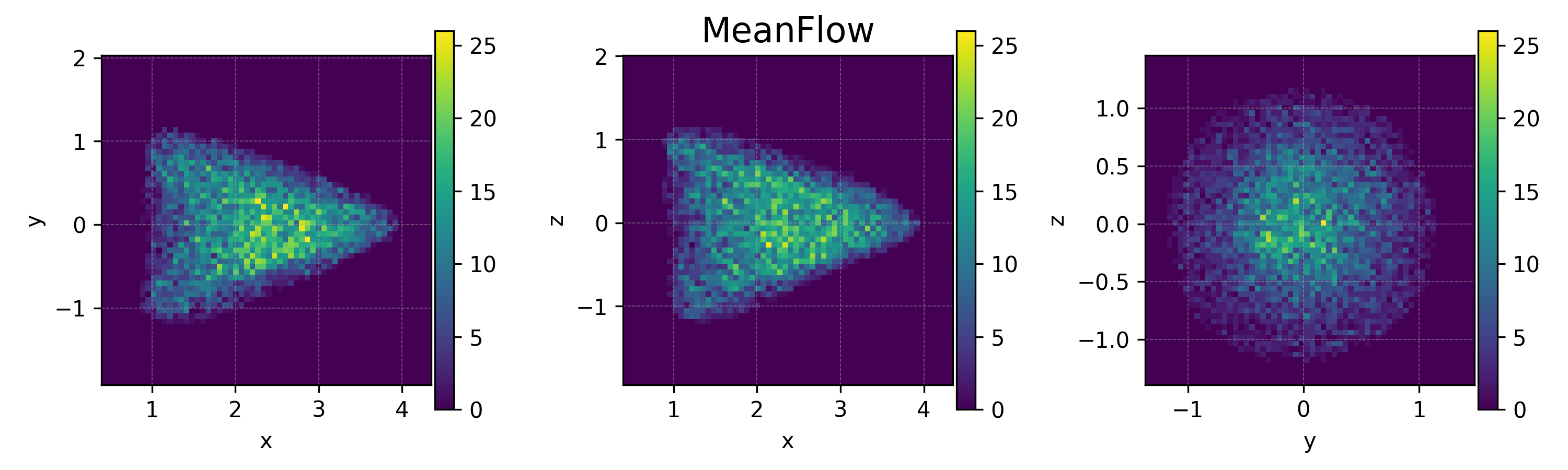}
  \end{subfigure}
  \begin{subfigure}{0.95\linewidth}
    \centering
    \includegraphics[width=\linewidth]{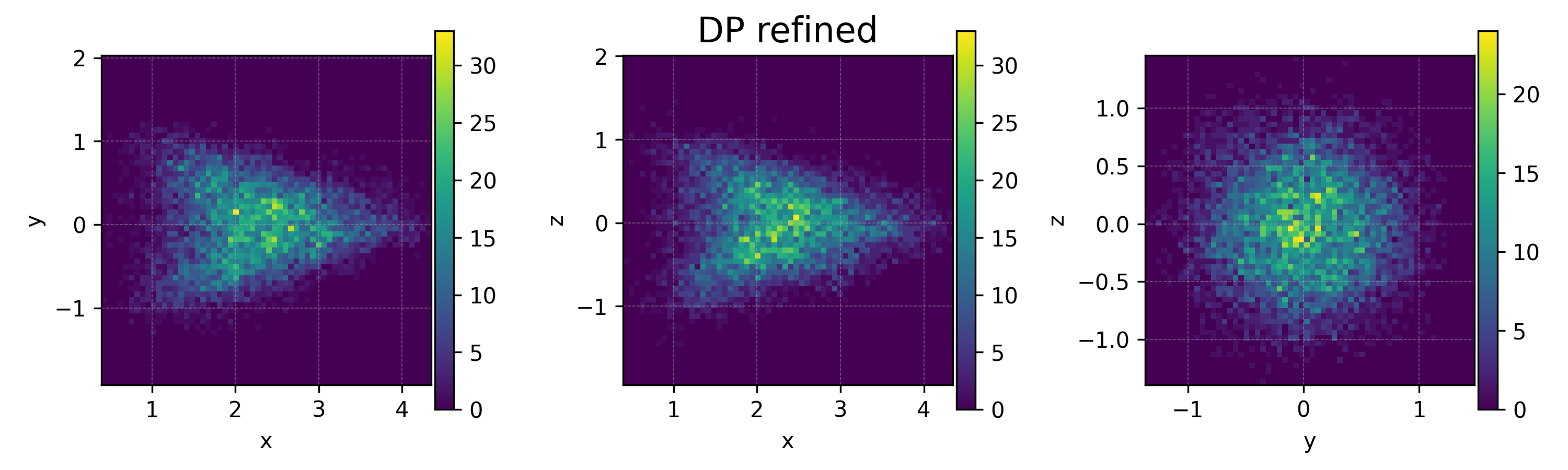}
  \end{subfigure}
  \caption{Qualitative comparisons on x-y, x-z, y-z at $\sigma=160$ for the 3D Keller-Segel system in a laminar flow. (a) Reference solution projected to three coordinate planes. (b) Predicted solution projected to the three coordinate planes by Meanflow ($W_2=0.0403$) (c) DP Refinement solution projects to the three coordinate planes ($W_2=0.0082$). }
  \label{fig:triptych180}
\end{figure}

\subsection{3D KS in Kolmogorov flow with Chaotic Streamlines}
\label{subsec:ksKol}
\begin{figure}[h]
  \centering
  \begin{subfigure}{0.9\linewidth}
    \centering
    \includegraphics[width=\linewidth]{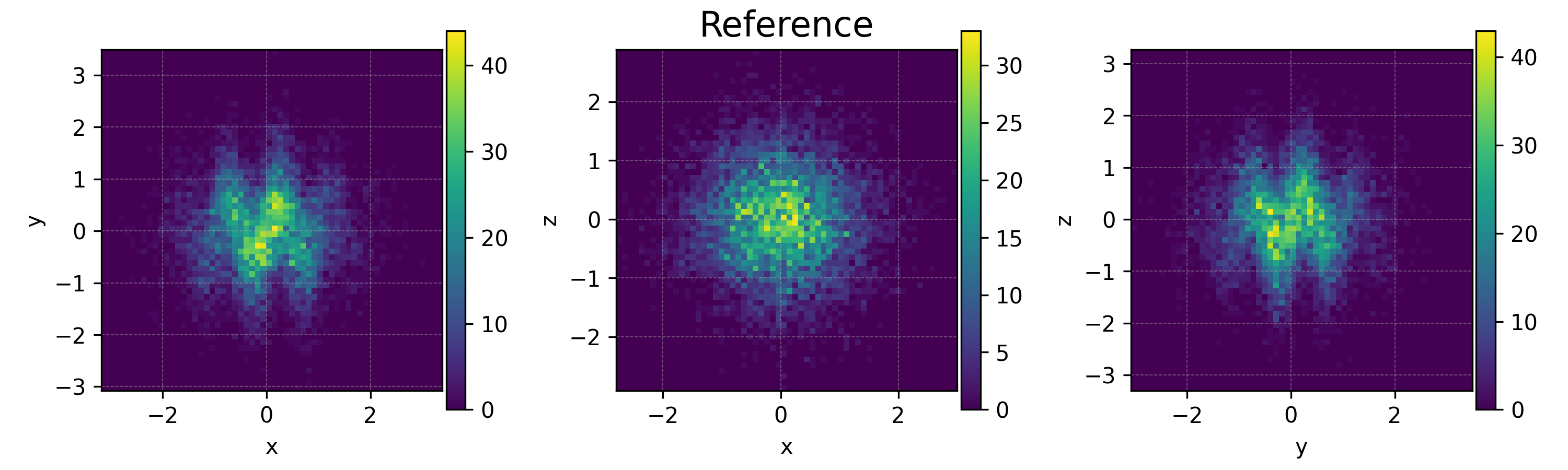}
   
  \end{subfigure}
  \begin{subfigure}{0.9\linewidth}
    \centering
    \includegraphics[width=\linewidth]{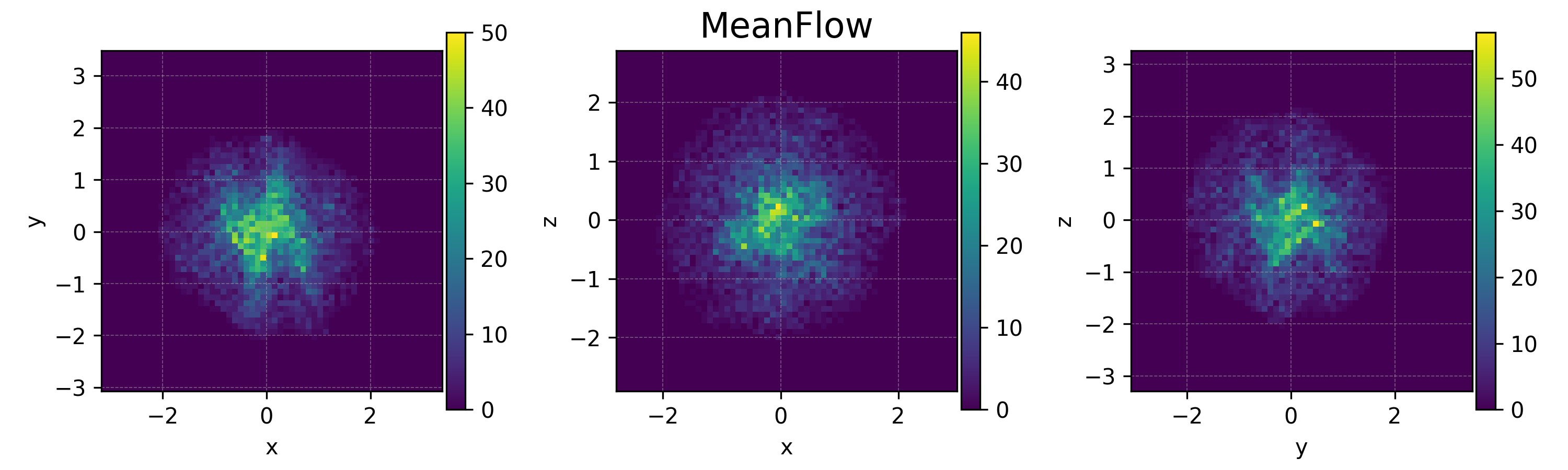}
    
  \end{subfigure}
  \begin{subfigure}{0.9\linewidth}
    \centering
    \includegraphics[width=\linewidth]{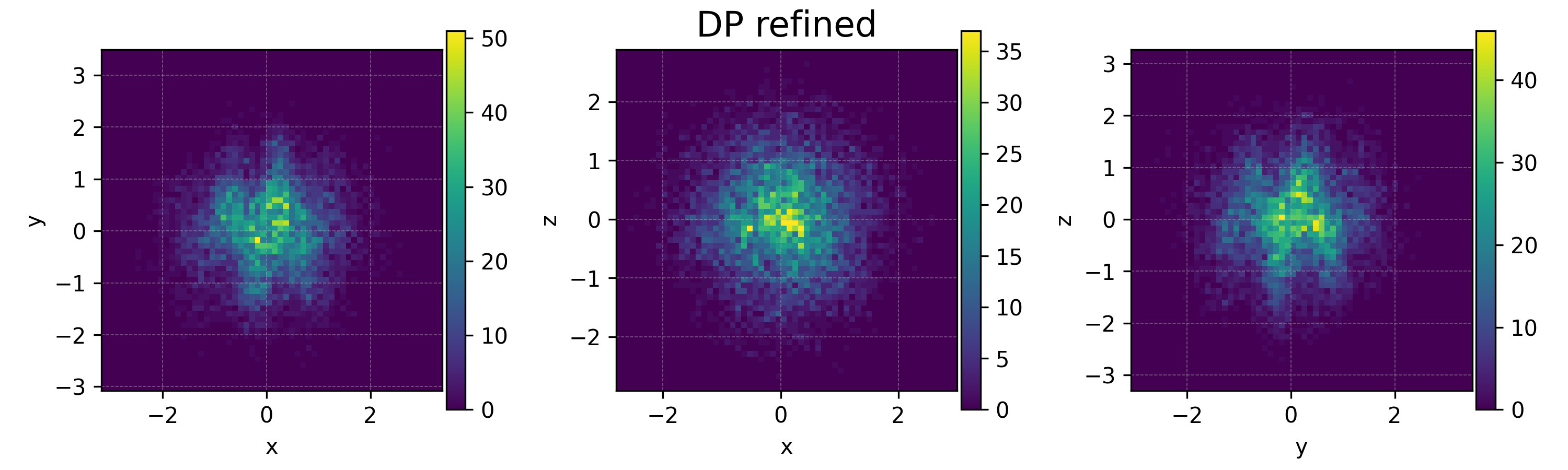}
    
  \end{subfigure}
  \caption{Qualitative comparisons of x-y, x-z, y-z  planes at $\sigma=110$ for the 3D Keller-Segel system in Kolmogorov flow. (a) Reference solution projected to three coordinate planes. (b) Predicted solution projected to the three coordinate planes by Meanflow (c) DP Refinement solution projects to the three coordinate planes.}  
  \label{fig:kol110}
\end{figure}

\noindent In this experiment, we study our method for the classical parabolic-elliptic 3D Keller-Segel system (or the reduced nonlinear and nonlocal $\rho$ equation~\ref{ks}) in the presence of Kolmogorov flow with chaotic streamlines \cite{CG95,BAMS_24}:
\begin{equation}
    \mathbf{v}(x,y,z) = \sigma \cdot 
\bigl( \sin(2\pi z),\, \sin(2\pi x),\, \sin(2\pi y) \bigr)^{T}.
\end{equation}
The data generation process is almost the same, except we use $n_{\text{dict}}{=}10$ amplitudes that are \emph{uniformly spaced} on $[10,100]$, defined by
\[
\sigma_k  =  10  +  \frac{k-1}{\,n_{\text{dict}}-1\,}\,(100-10),\qquad k=1,\dots,n_{\text{dict}}.
\]
The simulation results are shown in Figure~\ref{fig:kol110}. At $\sigma=110$, an extrapolation point beyond the training grid $\sigma\in\{10,\ldots,100\}$, the reference projections in Figure~\ref{fig:kol110} display the Kolmogorov-flow–induced anisotropy: narrow, high-density ridges aligned with coordinate directions in x-y and y-z, together with a more isotropic footprint in x-z. One–step Meanflow captures the gross concentration but remains over-smoothed, attenuating the ridge contrast and slightly biasing the barycenter. The DP corrector (near-identity, mini-batch OT) sharpens the directional structures: ridges become thinner and more pronounced in x-y and y-z while the x-z slice regains the compact, near-circular core. Crucially, this improvement at $\sigma{=}110$ (extrapolation) indicates that our two-step pipeline generalizes robustly out of distribution while preserving the 1-NFE determinism of MF.

\subsection{KPP Front Speed Experiment}
\label{subsec:kpp}
\noindent Front propagation in fluid flows becomes a hot topic in the physical sciences ~\cite{xin2009introduction}. The reaction-diffusion-advection equation with Kolmogorov-Petrovsky-Piskunov (KPP) nonlinearity is as follows~\cite{kolmogorov1937investigation}: 
\begin{equation}
  u_t + \mathbf{v}(x)\cdot\nabla u = \kappa \Delta u + r\,u(1-u),
\end{equation}
where $\kappa$ is diffusion constant, $\mathbf{v}$ is an incompressible velocity field , and $u$ is the concentration of reactant. If the velocity field is $T$-periodic in space and time, the minimal front speed satisfies the variational formula~\cite{nolen2005existence}:
\begin{equation}
  c^*(e) = \inf_{\lambda>0}\frac{\mu(\lambda)}{\lambda}
\end{equation}
where $\mu(\lambda)$ is the principal eigenvalue of parabolic operator $\partial_t - \mathcal{A}$ with:
\begin{equation}
    \mathcal{A}w
  := \kappa \Delta_x w
   + \big(2\kappa\lambda\,\mathbf e_1 + \mathbf v\big) \cdot \nabla_x w
   + \big(\kappa\lambda^{2} + \lambda\,\mathbf v \cdot \mathbf e_1 + 1\big)\,w
    \label{eq1}
\end{equation}
For a fixed $\lambda>0$, the twisted diffusion on $\mathbb{T}^2=[0,2\pi)^2$ is
\begin{equation}
  \mathrm{d}X_t
  = \bigl( 2\kappa\lambda\,\mathbf{e}_1 + \mathbf{v}(X_t,t) \bigr)\,\mathrm{d}t
    + \sqrt{2\kappa}\,\mathrm{d}W_t
  \label{eq:kpp-sde}
\end{equation}
with
\begin{equation}
    \mathbf{v}(x,t)=
  \begin{pmatrix}
    -\cos(x_2)-\theta\cos(2\pi t)\,\sin(x_1)\\
    \phantom{-}\cos(x_1)+\theta\cos(2\pi t)\,\sin(x_2)
  \end{pmatrix} ,
\end{equation}
which corresponds to the operator in Equation~\ref{eq1}.  The Feynman--Kac semigroup uses the potential function
\begin{equation}
     V_\lambda(x,t)=\kappa\lambda^2+\lambda\,\mathbf{v}(x,t) \cdot \mathbf{e}_1.
\end{equation}
With one forcing period normalized to $ \Delta T=1$  and each period discretized
into $r_n=2^{-\log_2\Delta t}$ substeps of size $\Delta t=1/r_n$, we first do the Euler--Maruyama propagation of \eqref{eq:kpp-sde}, then weighting multiplicatively by
\begin{equation}
w_j^i=\exp\big(V_\lambda(X_j^i,t_j)\Delta t\big)
\end{equation}
 and finally carrying out a multinomial resampling $\propto\{w_j^i\}$ in each substep to realize the potential effects and control variance in particle evolution.  The running
estimators are
\begin{equation}
  \widehat{\mu}_T(\lambda)
   =  \kappa + \frac{1}{T}\sum_{t=1}^{T r_n}\log\overline{w}_t,
  \qquad
  \widehat{c}_T(\lambda)=\widehat{\mu}_T(\lambda)/\lambda,
  \label{eq:mu-estimator}
\end{equation}
where $\overline{w}_j:=\tfrac{1}{M}\sum_i w_j^i$.
Meanwhile, let $\gamma_n(\varphi)=\mathbb{E}\left[\varphi(X_n)\prod_{t=1}^n\overline{w}_t\right]$
and $\eta_n=\gamma_n/\gamma_n(1)$ be the normalized empirical measure after resampling. Then $\eta_n \Rightarrow \eta^\star_\lambda$ and
$\tfrac{1}{n}\log\gamma_n(1)\to\mu(\lambda)$ as $n\to\infty$.
Consequently, we approximate $\eta^\star_\lambda$ by running the Feynman--Kac (FK) particle system long enough and collecting the terminal cloud. In this setting, we start
from the uniform distribution on $[0,2\pi)^2$ and uses a large $T=4096$.\\
In this experiment, We study KPP front–speed estimation on the two–torus $\mathbb{T}^2=[0,2\pi)^2$ under a time–periodic cellular flow.
We fix $(\lambda,\theta)=(2.0,1.0)$ and discretize each forcing period to $r_n=256$ substeps ($\log_2\Delta t=-8$).
Each run uses $M=2\times10^4$ particles; we print $\widehat{c}_T(\lambda)=\widehat{\mu}_T(\lambda)/\lambda$ at dyadic generations $T=1,2,4,\ldots$.
Two initializations are compared: \emph{warm} (Meanflow with DP refinement; Meanflow samples from a Gaussian source with std $\approx\pi$ and a 6-block residual corrector on the torus refines it, both conditioned on $\sigma$) and \emph{cold} (Uniform on $[0,2\pi)^2$). The $sigma$ is defined as the diffusion constant $\kappa$ in ~\ref{eq1} in this scenario.
We use dyadic diffusion constant levels $\sigma=2^{\ell}$ with uniformly spaced exponents for training 
\begin{equation}
  \ell^{\mathrm{tr}}_{k} = -2 - (k-1)\cdot 0.25, \qquad k=1,\ldots,8,
\end{equation}
and testing:
\begin{equation}
  \ell^{\mathrm{te}}_{k} = -2 - (k-1)\cdot 0.25, \qquad k=1,\ldots,11.
\end{equation}
with the two metrics:
\begin{enumerate}
  \item \textit{Eigenvalue convergence} of $\widehat{c}_T(\lambda)$, as defined in Equation~\ref{eq:mu-estimator}, versus 
  the number of FK generations $T \in \{1, 2, 4, \dots\}$;
  \item \textit{Invariant-measure accuracy:} For each $\sigma$, we quantify the invariant-measure error on the two-torus by
\begin{equation}\label{eq:inv_measure_2d}
\mathcal{E}(\sigma)
= W_2\!\left(\hat{\mu}^{(\sigma)},\,\hat{\nu}^{(\sigma)}\right),
\end{equation}
where $\hat{\mu}^{(\sigma)}$ and $\hat{\nu}^{(\sigma)}$ denote the empirical measures of the generated terminal particle cloud and the long-horizon FK reference particle cloud on $\mathbb{T}^2=[0,2\pi)^2$ at diffusion level $\sigma$, respectively. In practice, we evaluate \eqref{eq:inv_measure_2d} using a $K\times K$ histogram discretization on $[0,2\pi)^2$ and compute $W_2$ with the periodic ground metric on $\mathbb{T}^2$.
\end{enumerate}
\begin{figure}[H]
    \centering
    \begin{subfigure}{0.32\textwidth}
        \includegraphics[width=\linewidth]{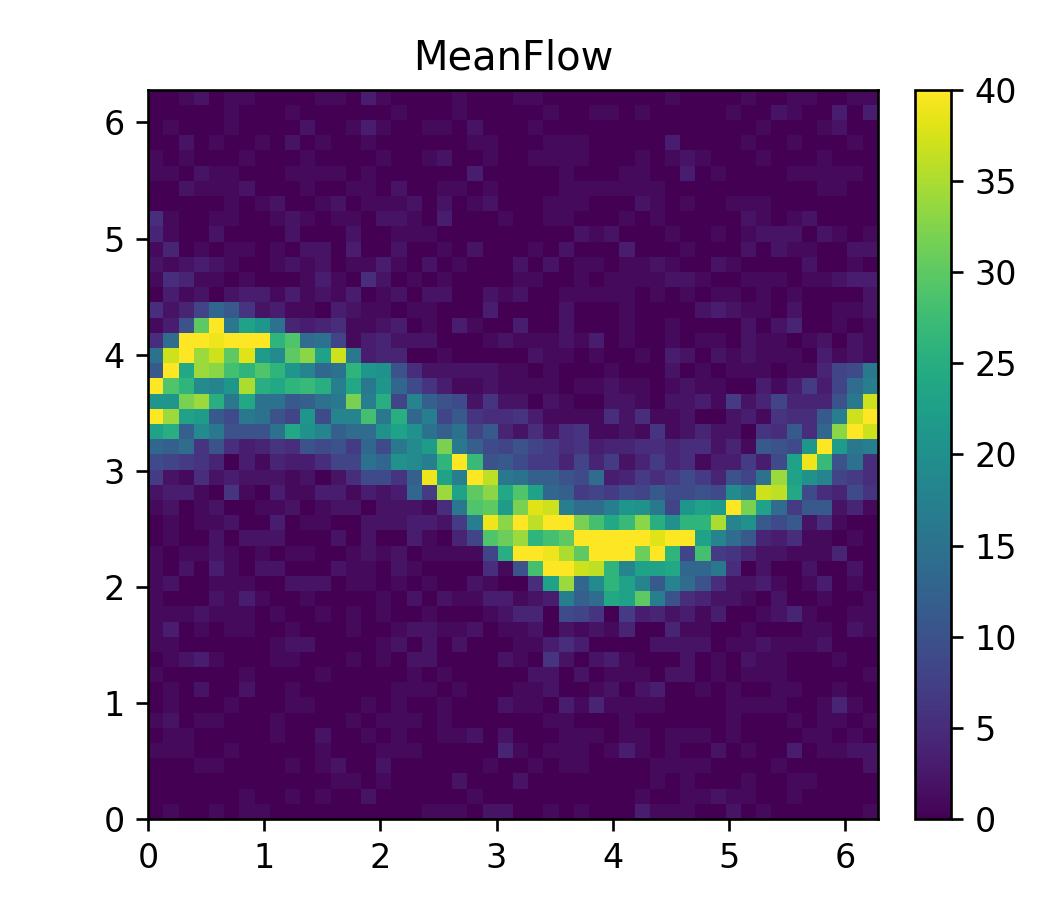}
    \end{subfigure}
    \hfill
    \begin{subfigure}{0.32\textwidth}
        \includegraphics[width=\linewidth]{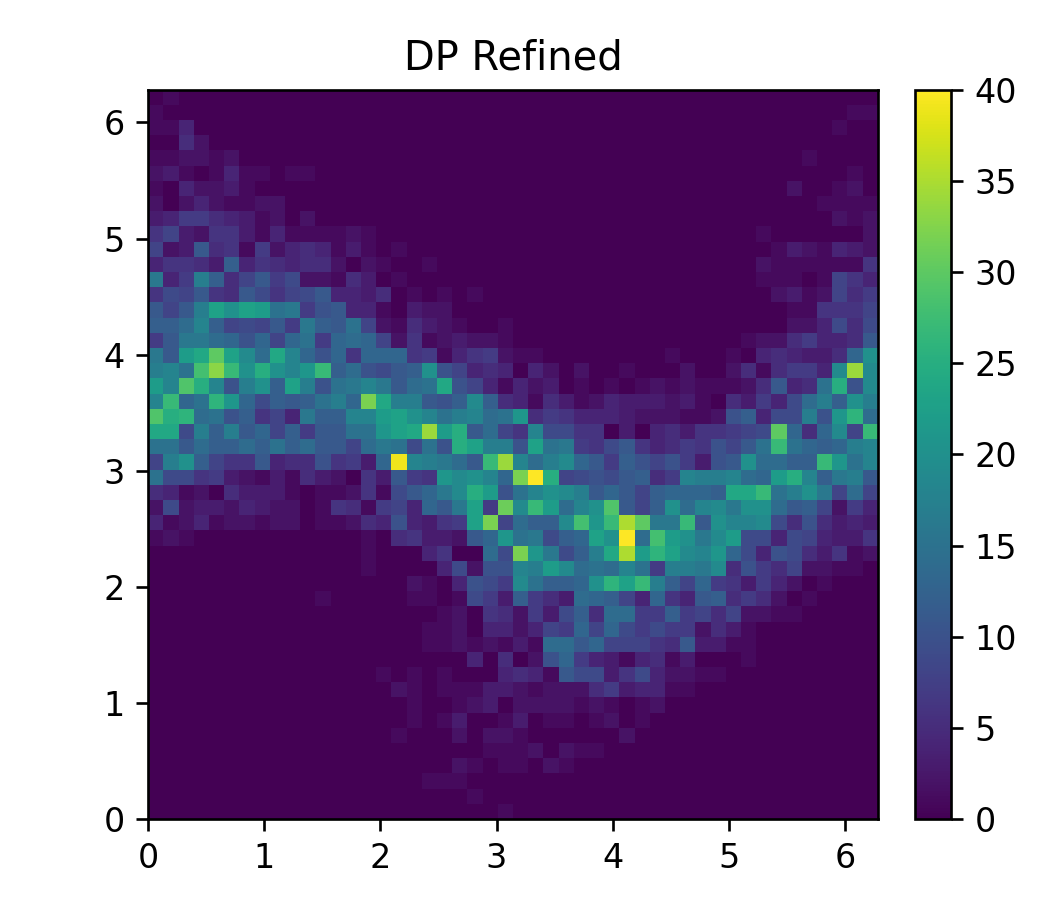}
    \end{subfigure}
    \hfill
    \begin{subfigure}{0.32\textwidth}
        \includegraphics[width=\linewidth]{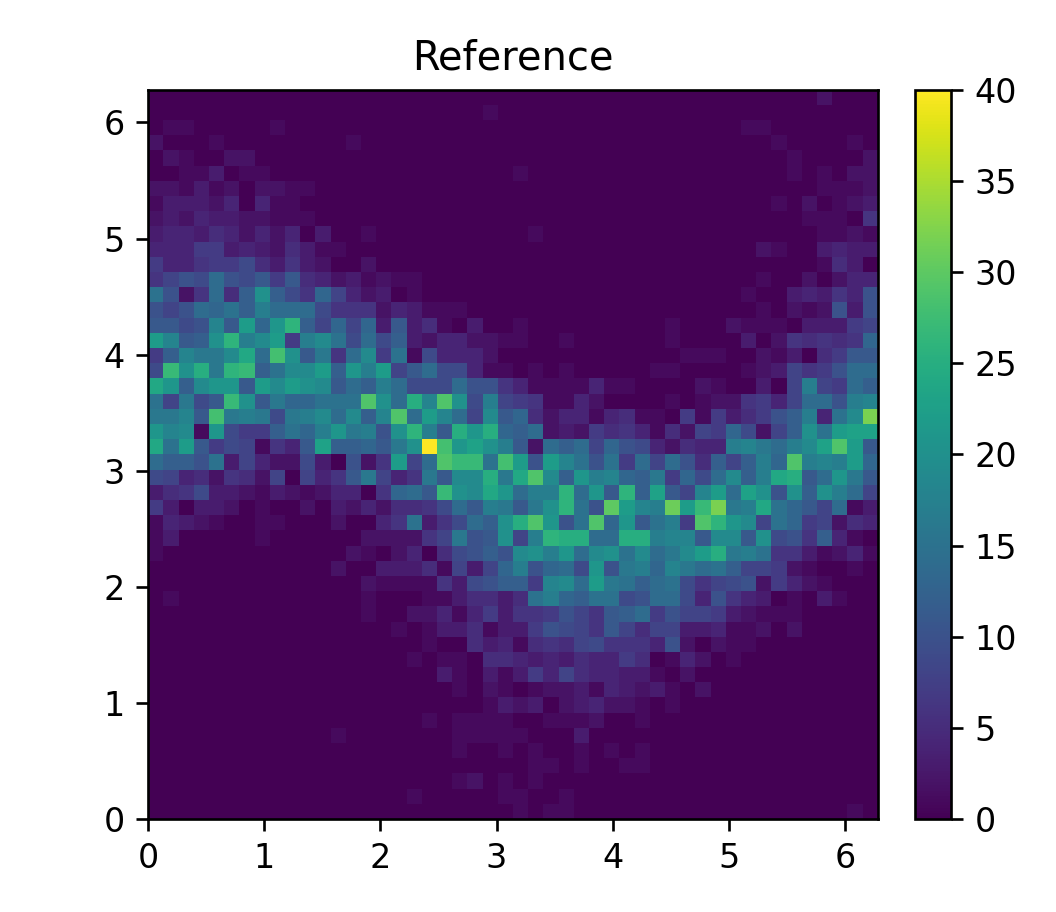}
    \end{subfigure}

    \vspace{1em} 

    \begin{subfigure}{0.32\textwidth}
        \includegraphics[width=\linewidth]{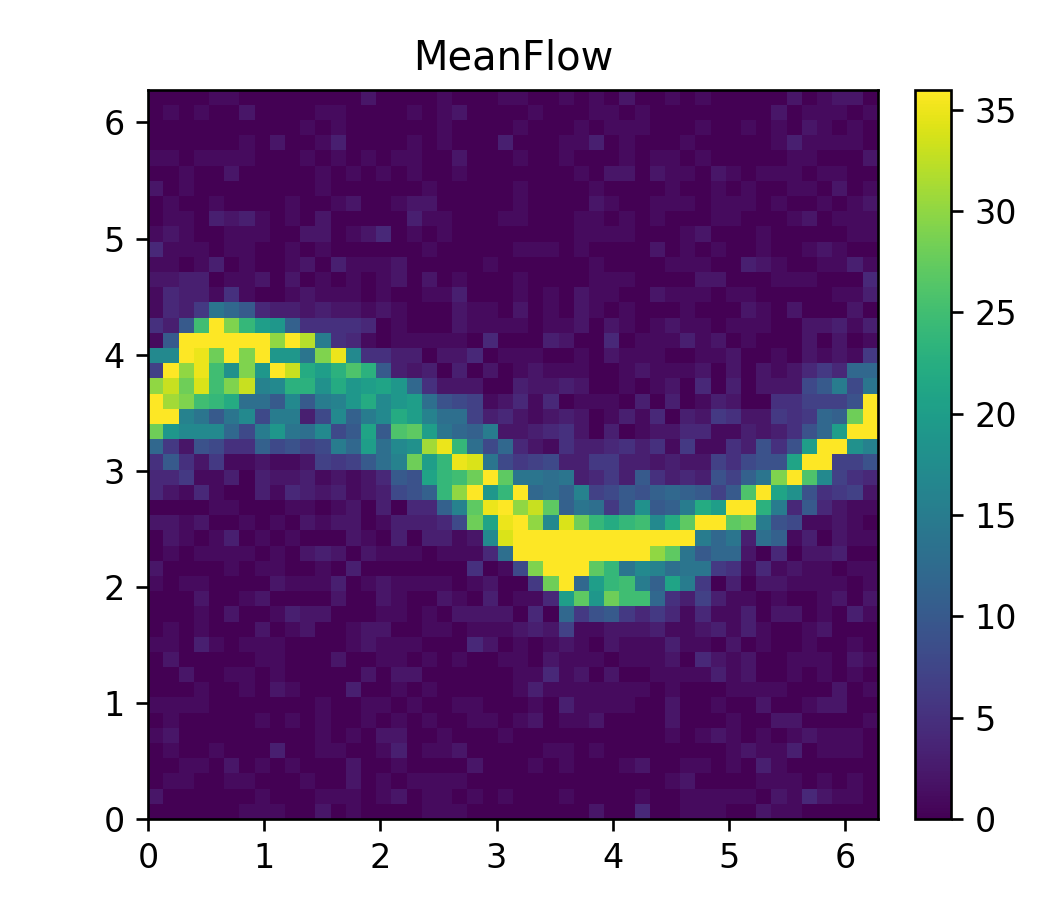}
    \end{subfigure}
    \hfill
    \begin{subfigure}{0.32\textwidth}
        \includegraphics[width=\linewidth]{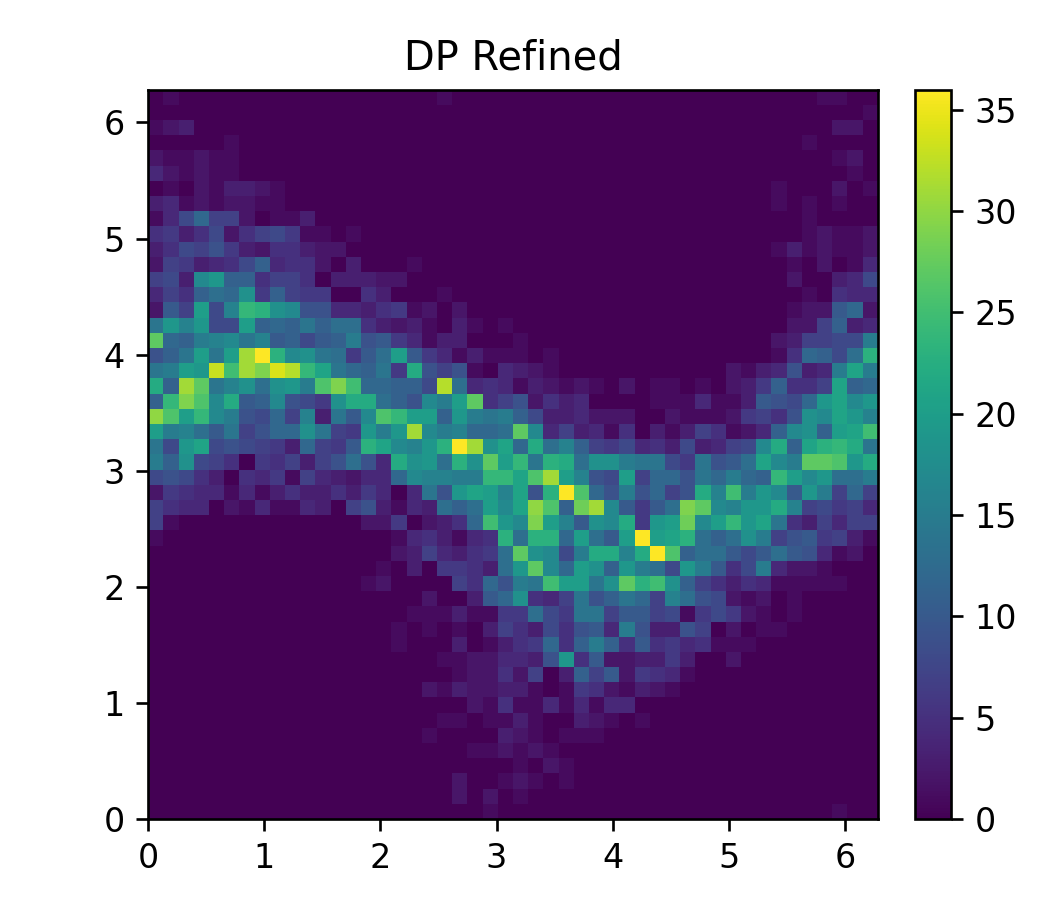}
    \end{subfigure}
    \hfill
    \begin{subfigure}{0.32\textwidth}
        \includegraphics[width=\linewidth]{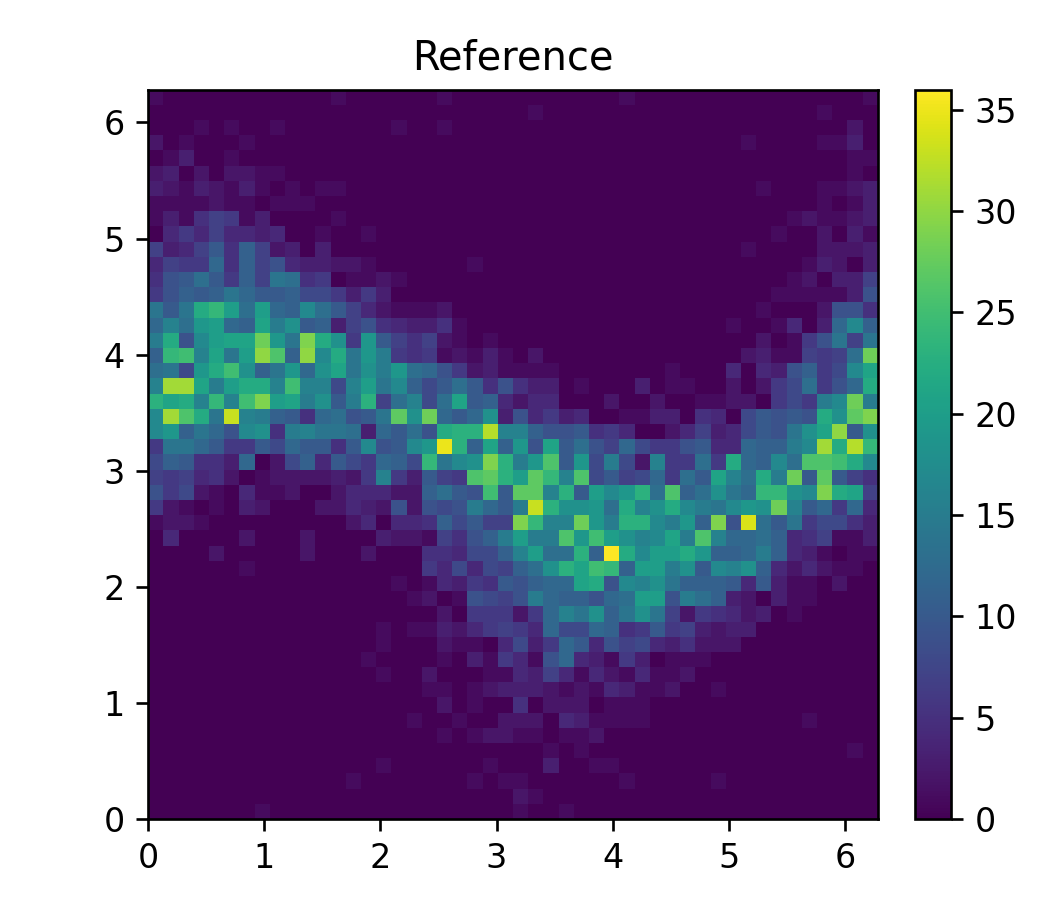}
    \end{subfigure}
    \caption{\textbf{Empirical invariant measures on $\mathbb{T}^2$ across diffusion constants.}
Rows correspond to $\sigma=2^{-3.75}$ (top, \emph{within the training range})
and $\sigma=2^{-4}$ (bottom, \emph{extrapolation beyond training});
columns show Meanflow (left), MF+DP (middle), and the resolved FK reference (right).
Each panel is a $72{\times}72$ histogram on $[0,2\pi)^2$.
The DP corrector contracts spurious mass and sharpens the anisotropic ridge, bringing the warm-start distribution visibly closer to the reference at both $\sigma$, including the extrapolation case.}

    \label{fig:kpp_measure}
\end{figure}
\begin{figure}[h]
    \centering
\begin{minipage}{0.35\linewidth}
\centering
\scriptsize
\begin{tabular}{lcc}
\toprule
\textbf{$-\log_2\sigma$} & \textbf{Meanflow} & \textbf{DP refinement} \\
\midrule
    2.00 & 0.0477 & \textbf{0.0138} \\
2.25 & 0.0476 & \textbf{0.0139} \\
2.50 & 0.0532 & \textbf{0.0124} \\
2.75 & 0.0592 & \textbf{0.0123} \\
3.00 & 0.0639 & \textbf{0.0106} \\
3.25 & 0.0717 & \textbf{0.0148} \\
3.50 & 0.0870 & \textbf{0.0200} \\
3.75 & 0.0773 & \textbf{0.0154} \\
4.00 & 0.1012 & \textbf{0.0216} \\
4.25 & 0.1195 & \textbf{0.0154} \\
4.50 & 0.1176 & \textbf{0.0456} \\
\bottomrule
\end{tabular}
\end{minipage}
\hfill
\begin{minipage}{0.56\linewidth}
\centering
\includegraphics[width=\linewidth]{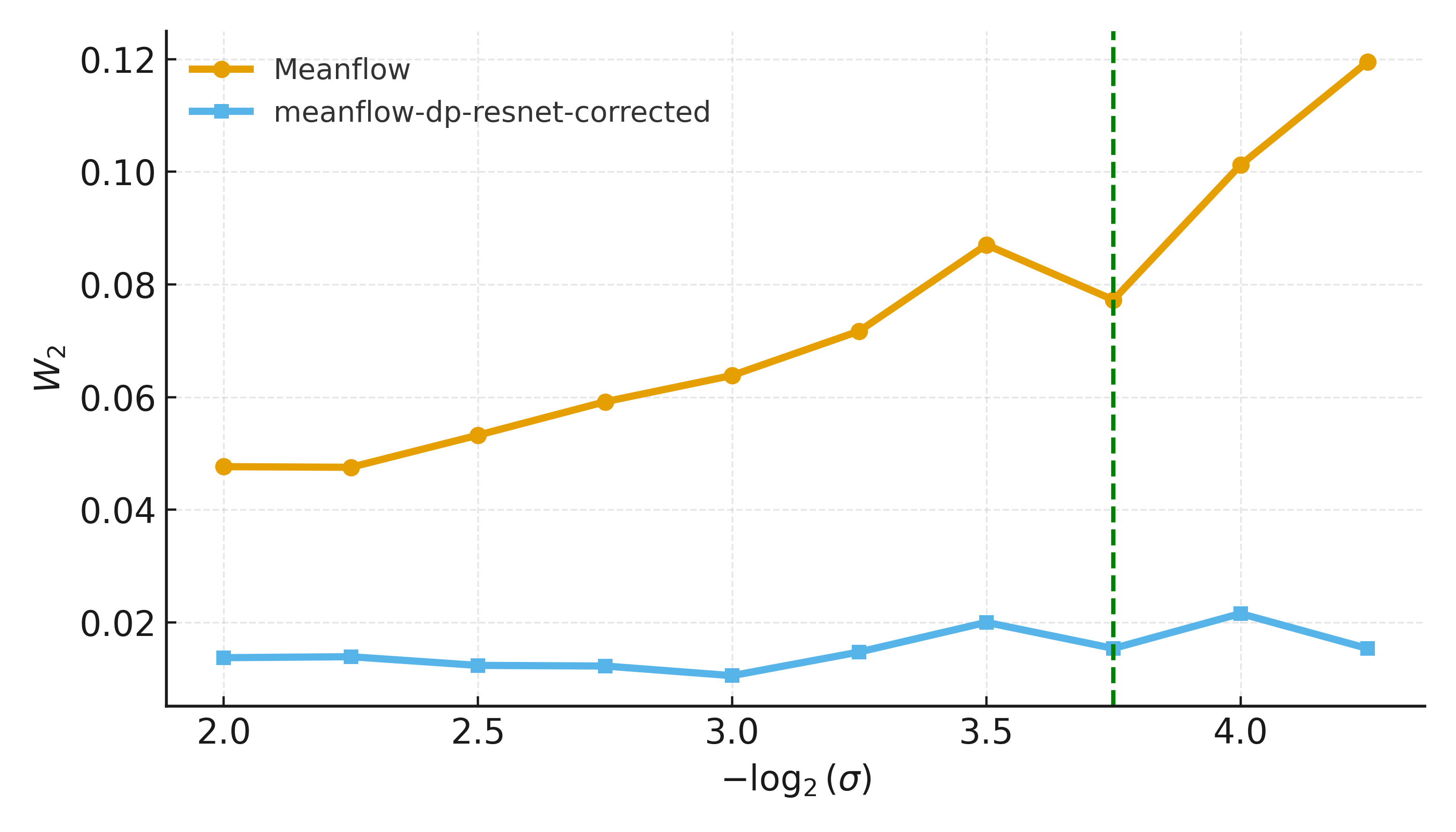}
\end{minipage}
    \caption{\emph{Left:} $W_2$ (lower is better) between the empirical measure and the long-$T$ reference for MF and the DP refinement, evaluated at $\sigma=2^{-\ell}$.
\emph{Right:} The same numbers plotted versus $-\log_2\sigma$; the green dashed line marks the end of the \emph{training} range ($-\log_2\sigma=3.75$), so points to the right are \emph{extrapolation}.
Across the entire grid, including the extrapolation regime, the DP refinement uniformly reduces error relative to MF and flattens the growth of $W_2$.}

\label{fig:w2_combokpp}
\end{figure}
\begin{figure}[h]
 \centering
\begin{minipage}{0.35\linewidth}
\centering
\scriptsize
\begin{tabular}{rcc}
\toprule
iteration & \textbf{MF-only init} & \textbf{MF+DP init}  \\
\midrule
1   & 0.733899 & 0.761573 \\
2   & 0.763092 & 0.773892 \\
4   & 0.770337 & 0.788397 \\
8   & 0.795811 & 0.781912 \\
16  & 0.784826 & 0.785587 \\
32  & 0.787654 & 0.783146 \\
64  & 0.787575 & 0.775075 \\
128 & 0.783591 & 0.779001 \\
\bottomrule
\end{tabular}
\end{minipage}
\hfill
\begin{minipage}{0.56\linewidth}
\centering
\includegraphics[width=\linewidth]{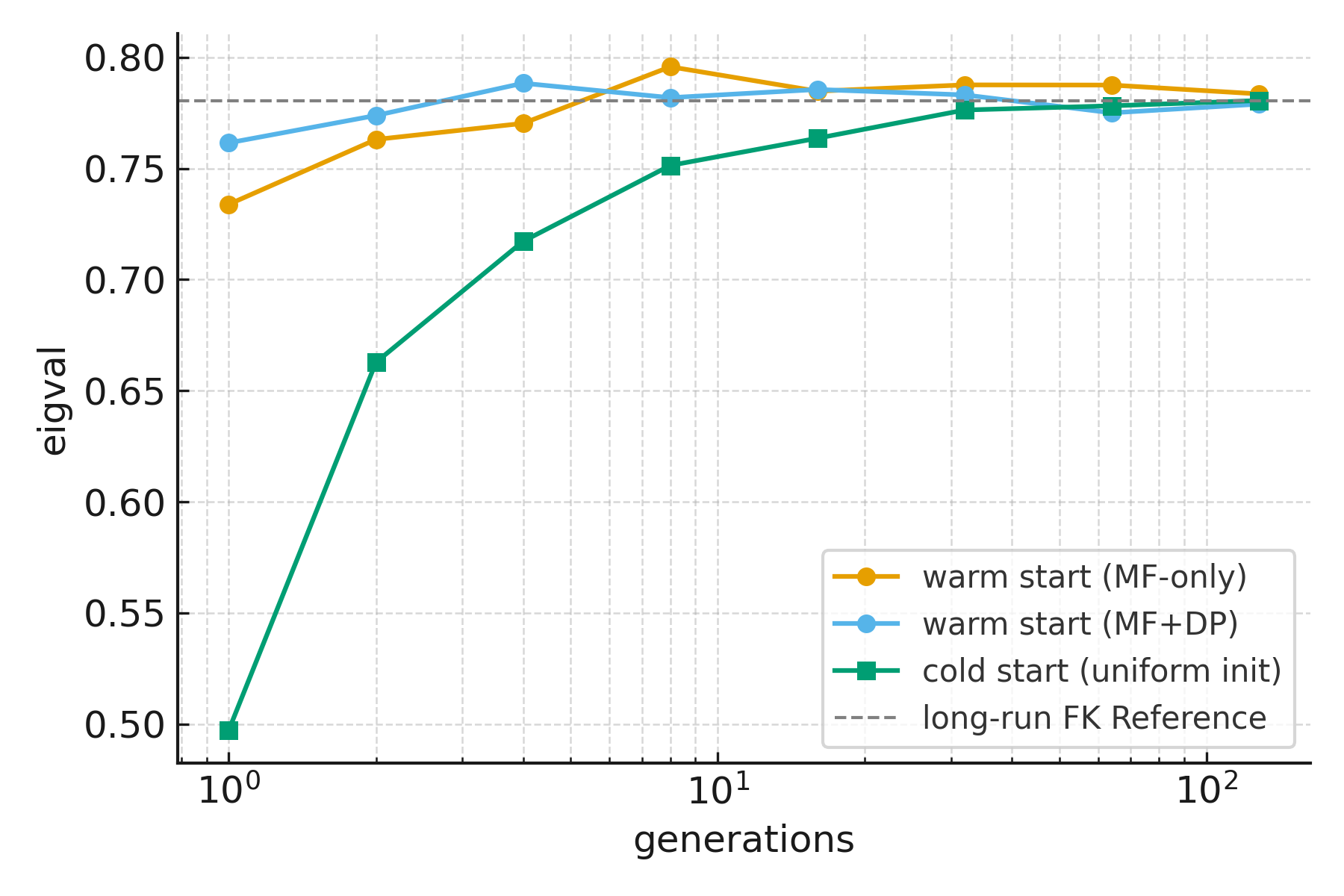}
\end{minipage}
  \caption{Eigenvalue convergence with warm and cold starts.
  The plot shows \(\hat c_T(\lambda)=\hat\mu_T(\lambda)/\lambda\) versus generations \(T\) (log scale) for
  warm (MF-only), warm (MF+DP), and cold (uniform) initializations. 
  The dashed line is the long-run FK reference.
  The left table reports the estimator at early iterations \(T\in\{1,2,4,\ldots,128\}\) for the two warm starts.}
  \label{fig:eig-conv}
\end{figure}
\noindent Figure~\ref{fig:kpp_measure} compares empirical invariant measures on $\mathbb{T}^2$ for two representative diffusion constants: $\sigma=2^{-3.75}$ (within training) and $\sigma=2^{-4}$ (extrapolation).
Across both rows the MF+DP (middle) corrects the one–shot Meanflow bias (left) by contracting spurious mass near the central trough and sharpening the anisotropic ridge that aligns with the cellular advection; the resulting density closely matches the long–run FK reference (right).
Notably, the same geometric improvement persists at $\sigma=2^{-4}$, evidencing strong out–of–range generalization.\\
Figure~\ref{fig:w2_combokpp} summarizes the 2-Wasserstein distance on the torus between the empirical invariant measure and the long-$T$ reference.
Across the full grid, including extrapolation beyond the training range, the DP refinement consistently reduces $W_2$ relative to Meanflow (right panel).
The Meanflow error increases noticeably past the training boundary at $-\log_2\sigma=3.75$, whereas MF+DP remains stable and significantly lower.
For instance, at the extrapolation point $-\log_2\sigma=4.25$, $W_2$ decreases from $0.11950$ (MF) to $0.01539$ (MF+DP).
The left table in Fig.~\ref{fig:w2_combokpp} reports consistent improvements across all tested $\sigma$, and the right plot visualizes the same trend, with the training range boundary marked by a dashed line.
\\
Figure~\ref{fig:eig-conv} reports $\widehat{c}_T(\lambda)=\widehat{\mu}_T(\lambda)/\lambda$ versus generations $T$ for the warm(MF+DP), warm(MF only), and cold (uniform) starts. With DP refinement, the very first estimate (\(T=1\)) is already almost identical to the long-run FK reference and remains close thereafter, indicating minimal transient bias and variance. The MF-only warm start begins slightly below the reference but converges within a few generations. In contrast, the cold start begins far from the limit and requires roughly an order of magnitude more generations to catch up. Overall, DP refinement provides the strongest warm start: its initial eigenvalue essentially equals the long-run FK reference, and it stabilizes fastest.

\subsection{3D KPP Front Speed Experiment with time-dependent Kolmogorov flow}
\label{subsec:3dkpp}
\noindent In this section, we study our method on the 3D KPP equation:
\begin{equation}
u_t + \mathbf{v}(\mathbf{x}) \cdot \nabla u = \kappa \Delta u + r\,u(1 - u),
\end{equation}
with a three-dimensional time-dependent Kolmogorov flow velocity field:
\begin{equation}
\mathbf{v}(\mathbf{x}, t) = 
\bigl(
\sin(x_3 + \sin(2\pi t)),
\sin(x_1 + \sin(2\pi t)),
\sin(x_2 + \sin(2\pi t))
\bigr)
\end{equation}

\noindent The setup parallels that of the 2D experiment. The spatial domain is 
$\mathbb{T}^3 = [0, 2\pi)^3$, with the time period normalized to $T = 1$. We employ Euler--Maruyama integration with $\Delta t = 2^{-8}$ and perform FK propagation using $2 \times 10^4$ particles per diffusion level. The diffusion constant $\sigma$ (identical to $\kappa$) follows a dyadic grid $\sigma = 2^{-\ell}$. Training levels are defined for $-\log_2 \sigma \le 3.75$, and evaluation extends up to $-\log_2 \sigma = 4.5$ to test out-of-range generalization. Furthermore, we compare three initialization modes, where the first is the Cold start, where we start from the uniform distribution on $\mathbb{T}^3$ and this case served as the reference group. Meanwhile, we tested the two warm start methods through Meanflow and our two-step diffusion pipeline, where we would like to test our model on the following two metrics:
\begin{figure}[H]
  \centering
  \begin{subfigure}{\linewidth}
    \centering
    \includegraphics[width=\linewidth]{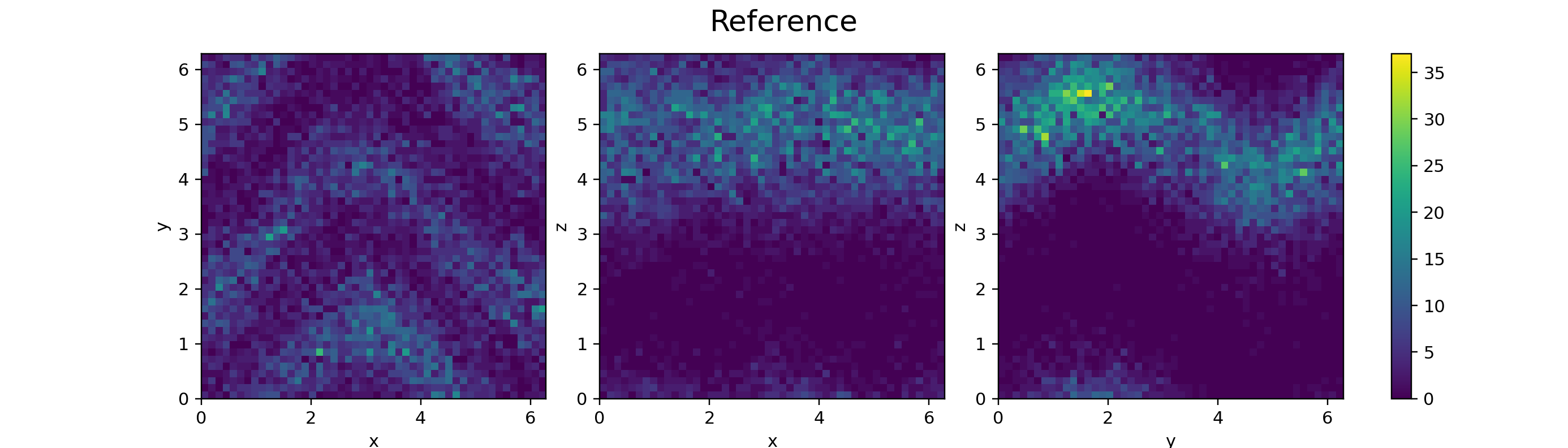}
   
  \end{subfigure}
  \begin{subfigure}{\linewidth}
    \centering
    \includegraphics[width=\linewidth]{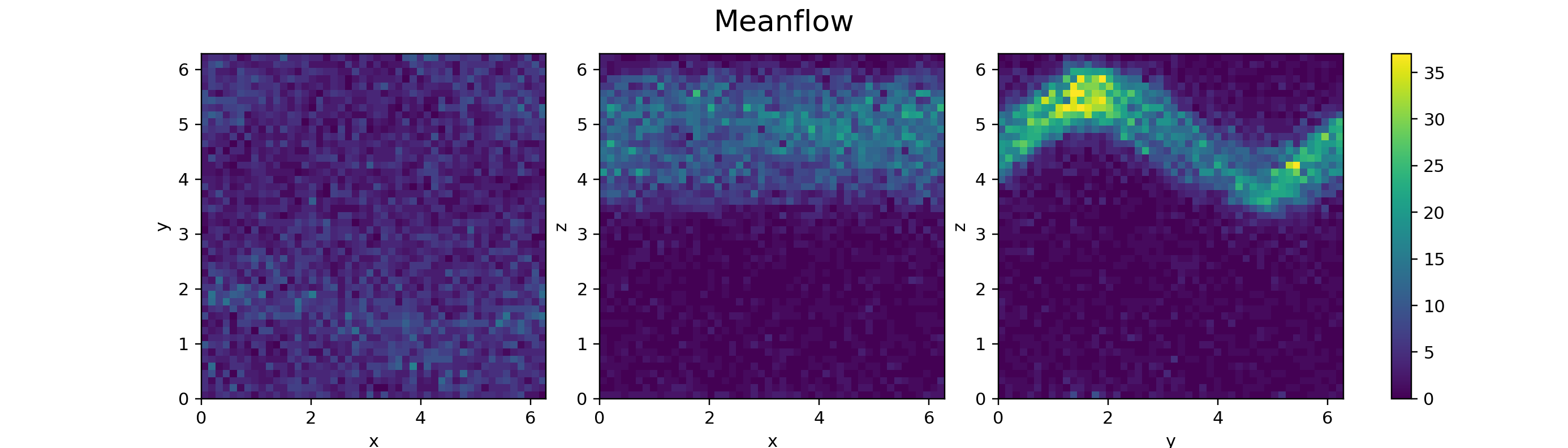}
    
  \end{subfigure}
  \begin{subfigure}{\linewidth}
    \centering
    \includegraphics[width=\linewidth]{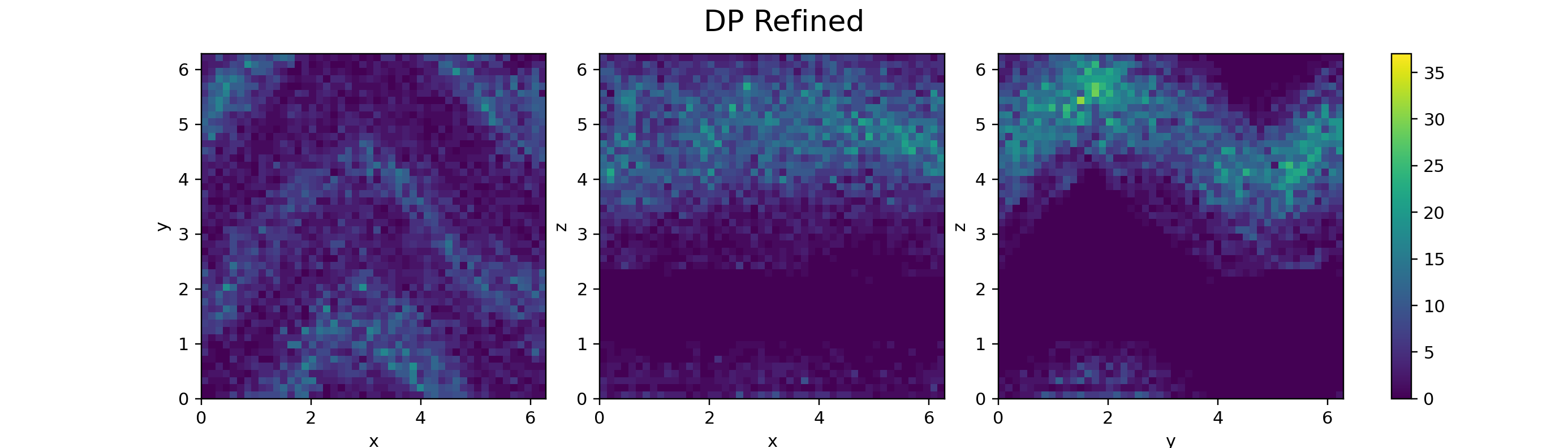}
    
  \end{subfigure}
  \caption{Qualitative comparisons of x-y, x-z, y-z  planes at $\sigma=2^{-4.5}$ for the 3D KPP system in 3D time-dependent Kolmogorov flow. (a) Reference solution projected to three coordinate planes. (b) Predicted solution projected to the three coordinate planes by Meanflow (c) DP Refinement solution projects to the three coordinate planes.}  
  \label{3dmeasure}
\end{figure}
\begin{enumerate}
  \item \textit{Eigenvalue convergence} of $\widehat{c}_T(\lambda)$, as defined in Equation~\ref{eq:mu-estimator}, versus 
  the number of FK generations $T \in \{1, 2, 4, \dots\}$;
  \item \textit{Invariant-measure accuracy:} For each $\sigma$, we quantify the invariant-measure error by
\begin{equation}
\mathcal{E}_{xz}(\sigma)
=
W_2\!\left(
(\Pi_{xz})_{\#}\widehat{\mu}^{(\sigma)},
(\Pi_{xz})_{\#}\widehat{\nu}^{(\sigma)}
\right),
\label{eq:invariant_measure_w2}
\end{equation}
where the projection operator $\Pi_{xz}:\mathbb{R}^3\to\mathbb{R}^2$ is defined as
$\Pi_{xz}(x,y,z)=(x,z)$, and $\widehat{\mu}^{(\sigma)}$ and
$\widehat{\nu}^{(\sigma)}$ denote the empirical measures of the generated
particle cloud and the FK reference particle cloud at parameter $\sigma$,
respectively.
\end{enumerate}
\begin{figure}[H]
    \centering
\begin{minipage}{0.33\linewidth}
\centering
\scriptsize
\begin{tabular}{lcc}
\toprule
\textbf{$-\log_2\sigma$} & \textbf{Meanflow} & \textbf{DP refinement} \\
\midrule
2.00 & 0.1591 & \textbf{0.0160} \\
2.25 & 0.1257 & \textbf{0.0230} \\
2.50 & 0.1442 & \textbf{0.0193} \\
2.75 & 0.1646 & \textbf{0.0280} \\
3.00 & 0.1250 & \textbf{0.0365} \\
3.25 & 0.1175 & \textbf{0.0314} \\
3.50 & 0.1342 & \textbf{0.0340} \\
3.75 & 0.1261 & \textbf{0.0253} \\
4.00 & 0.1049 & \textbf{0.0606} \\
4.25 & 0.1325 & \textbf{0.0282} \\
4.50 & 0.1099 & \textbf{0.0210} \\
\bottomrule
\end{tabular}
\end{minipage}
\hfill
\begin{minipage}{0.55\linewidth}
\centering
\includegraphics[width=\linewidth]{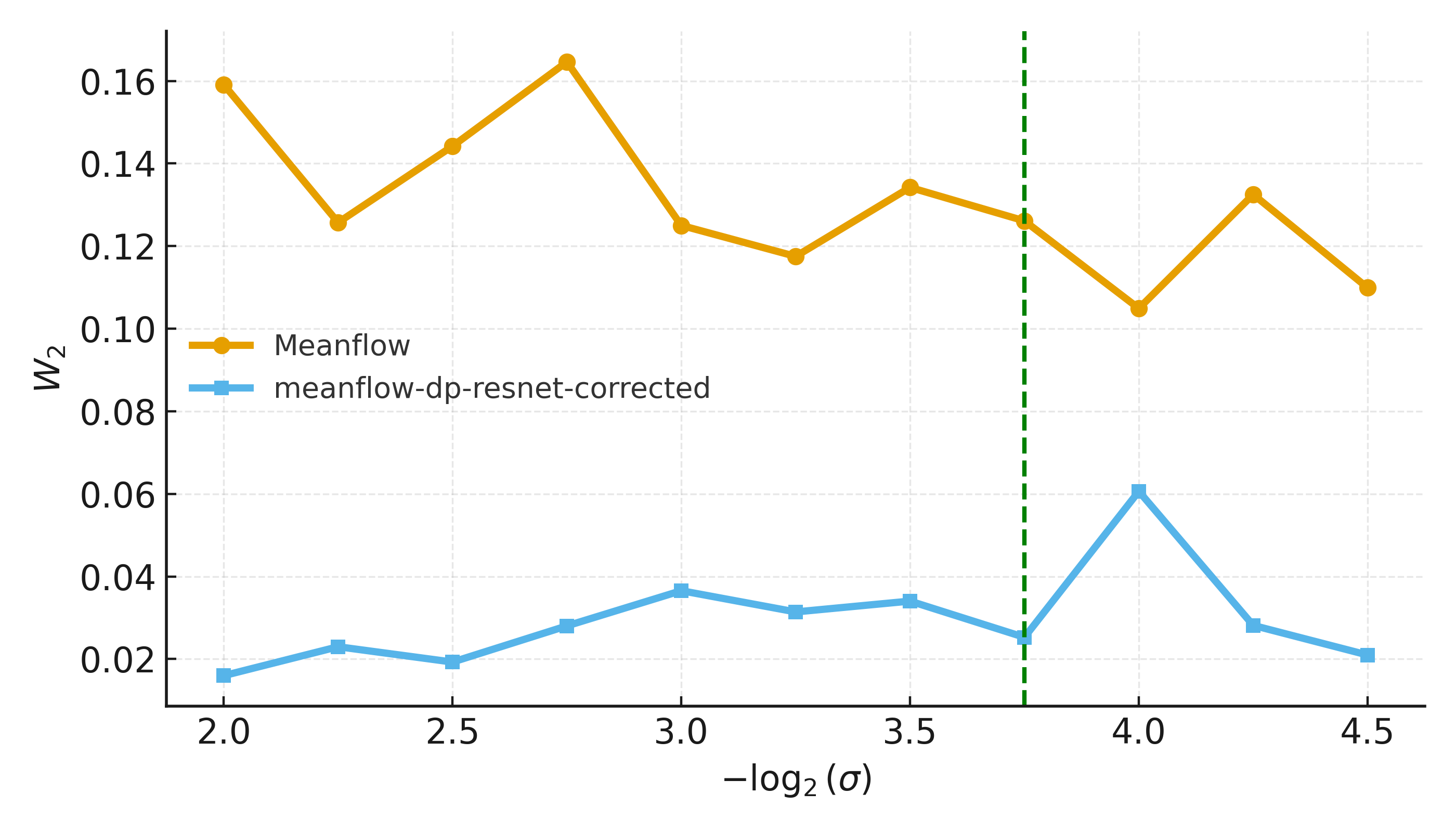}
\end{minipage}
    \caption{\emph{Left:} $W_2$ (Projection on the x-z plane) between the empirical measure and the long-$T$ reference for MF and the DP refinement, evaluated at $\sigma=2^{-\ell}$.
\emph{Right:} The same numbers plotted versus $-\log_2\sigma$; the green dashed line marks the end of the \emph{training} range ($-\log_2\sigma=3.75$), so points to the right are \emph{extrapolation}.
Across the entire grid, including the extrapolation regime, the DP refinement uniformly reduces error relative to MF and flattens the growth of $W_2$.}
\label{3dw2}
\end{figure}
\begin{figure}[H]
  \centering
\begin{minipage}{0.35\linewidth}
\centering
\scriptsize
\begin{tabular}{rcc}
\toprule
iteration & \textbf{MF-only init} & \textbf{MF+DP init} \\
\midrule
1   & 0.566547 & 0.613531 \\
2   & 0.551988 & 0.610457 \\
4   & 0.546482 & 0.621819 \\
8   & 0.576172 & 0.599699 \\
16  & 0.597887 & 0.590171 \\
32  & 0.609560 & 0.614999 \\
64  & 0.609690 & 0.615682 \\
128 & 0.614481 & 0.613865 \\
\bottomrule
\end{tabular}
\end{minipage}
\hfill
\begin{minipage}{0.56\linewidth}
\centering
\includegraphics[width=\linewidth]{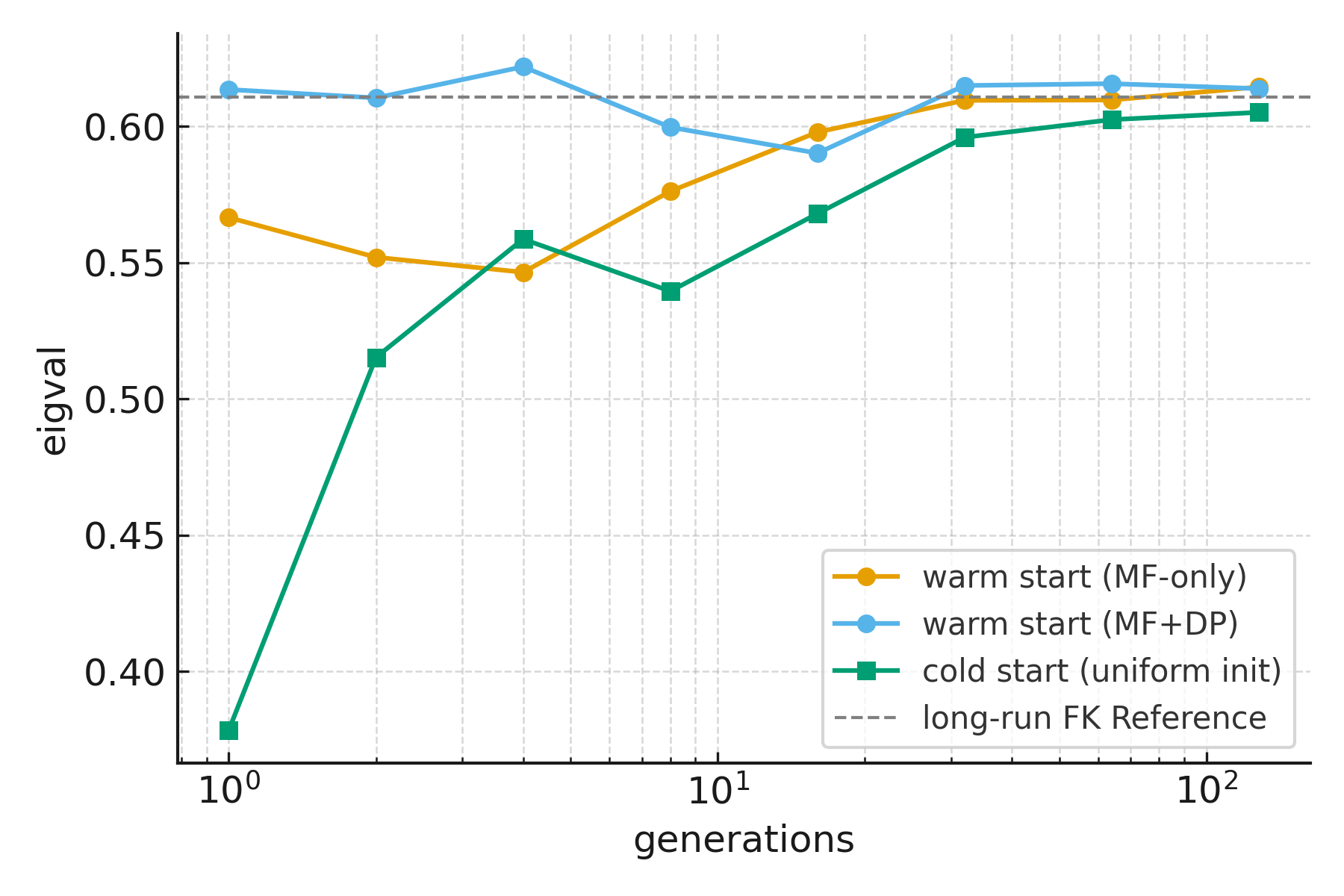}
\end{minipage}
  \caption{Eigenvalue convergence with warm and cold starts.
  The plot shows \(\hat c_T(\lambda)=\hat\mu_T(\lambda)/\lambda\) versus generations \(T\) (log scale) for
  warm (MF-only), warm (MF+DP), and cold (uniform) initializations. 
  The dashed line is the long-run FK reference.
  The left table reports the estimator at early iterations \(T\in\{1,2,4,\ldots,128\}\) for the two warm starts.}
  \label{fig:eig-conv3d}
\end{figure}
For the invariant measures, Figure~\ref{3dmeasure} shows triptych projections at
$\sigma = 2^{-4.5}$ for the FK reference (top row), Meanflow (middle), and
DP-refined outputs (bottom). Across the x-y, x-z, y-z planes, Meanflow exhibits spurious mass and blurred anisotropic ridges, whereas DP refinement removes these artifacts and sharpens the cellular alignment. In the x-y projection, DP restores the bifurcated high-density ridge; in x-z, it corrects the layer position and suppresses over-diffusion; and in y-z, it aligns the curved ridge with the FK reference.\\
Meanwhile, our pipeline achieves strong quantitative accuracy across diffusion levels. Figure~\ref{3dw2} reports the $W_2$ errors over $-\log_2\sigma\in[2.0,4.5]$.
Within the training range, the DP refinement consistently improves over Meanflow.
Beyond the training range, the method maintains stable extrapolation behavior, with substantially lower $W_2$ than the MF baseline. As an example, at $ \log_2\sigma=4.50$, $W_2$ decreases from $0.10990$ (MF) to $0.02098$ (DP), indicating effective out-of-range generalization.
\\
Last but not least, our model exhibits rapid eigenvalue convergence.  
Figure~\ref{fig:eig-conv3d} (and its accompanying table) reports
$\widehat{c}_T(\lambda)$ versus the number of FK generations. The
\emph{Warm (MF+DP)} initialization is nearly unbiased already at $T=1$ and then
remains essentially on the FK reference. The \emph{Warm (MF-only)} initialization
begins below the reference but converges within roughly $32$ generations,
demonstrating the stronger and more stable warm start provided by our two-step
diffusion pipeline compared with Meanflow alone.

\section{Conclusion}
\noindent We introduced a Two-Step Diffusion framework for fast and reliable generation in
Keller--Segel chemotaxis and Kolmogorov–Petrovsky–Piskunov (KPP) systems with background flows. As illustrated in Figure~\ref{fig:flow_chart}, our method separates
coarse global transport from fine local alignment: a Meanflow-style regressor as shown in Algorithm~\ref{alg:meanflow_train},
trained via the MeanFlow identity, provides a deterministic, 1-NFE initializer
that moves particles close to the correct support, and a near-identity
Deep Particle corrector, as shown in Algorithm~\ref{alg:dp_refine} then minimizes a mini-batch 2-Wasserstein objective
with exact EMD couplings. This coarse-to-fine design restores an explicit
geometry-aware $W_2$ training signal where it is tractable, making high-dimensional
OT optimization stable while preserving one-shot sampling.\\
Across 3D Keller--Segel systems with both laminar and Kolmogorov flows, the two-stage pipeline consistently lowers empirical $W_2$ (Figure~\ref{fig:w2_combo})relative to Meanflow and
sharpens anisotropic structure (Figure~\ref{fig:triptych180} and~\ref{fig:kol110}), with the largest gains in the
singular-perturbation and extrapolation regimes. The Appendix further shows that
these benefits strengthen in a 4D KS extension (Figure~\ref{fig:4d_ks_4d}), are robust under changes in
mini-batch size and coupling refresh rate~\ref{tab1}, and persist when comparing with
self-distilled IMM~\cite{zhou2025inductive,boffi2025build} and 
regularized
Sinkhorn solvers. In KPP front-speed experiments
on both 2D and 3D time-dependent flows, the warm start yields invariant measures substantially closer to long-horizon Feynman--Kac references (Figure~\ref{fig:kpp_measure} and~\ref{3dmeasure}) and eigenvalue estimates that are nearly unbiased from the first generation (Figure~\ref{fig:eig-conv} and~\ref{fig:eig-conv3d}),
highlighting Two-Step Diffusion as a practical recipe for combining fast one-step flows with principled OT-based refinement in scientific machine learning.

\section{Acknowledgements}
\noindent ZW was partly supported by NTU SUG-023162-00001, MOE AcRF
Tier 1 Grant RG17/24. JX was partly supported by NSF grants DMS-2219904 and DMS-2309520, a Qualcomm Gift Award; and the Swedish Research Council grant no. 2021-06594 at the Institut Mittag-Leffler in Djursholm, Sweden, and the E. Schr\"odinger Institute, Vienna, Austria, both during his stay in the Fall of 2025. ZZ was supported by the National Natural Science Foundation of China (Projects 92470103 and 12171406), the Hong Kong RGC grant (Projects 17304324 and 17300325), the Seed Funding Programme for Basic Research (HKU), the Outstanding Young Researcher Award of HKU (2020–21), and the Seed Funding for Strategic Interdisciplinary Research Scheme 2021/22 (HKU).
\appendix
\appendix
\section{Appendix}
\noindent This appendix provides additional experiments and analyses that complement the
main text. Section~\ref{subsec:4d_ks_extension} extends the Keller--Segel setup to a 4D laminar flow and
shows that the advantage of our two-stage design becomes even more pronounced in
higher dimensions. Section~\ref{sec:ablation} presents an ablation over the DP mini-batch size
and coupling refresh rate, demonstrating that our refinement is both efficient
and robust to these hyperparameters. Section~\ref{sec:comparison-baselines} compares Two-Step Diffusion
with IMM-style self-distilled one-step baseline 
while Section~\ref{sec:sinkhorn} contrasts exact mini-batch EMD with entropically regularized
Sinkhorn solvers. 
\subsection{4D Keller--Segel System Extensions}
\label{subsec:4d_ks_extension}
\noindent To assess whether our two--stage pipeline scales beyond three spatial dimensions, 
we extend the Keller--Segel dynamics in~\eqref{ks} to a four--dimensional
state variable \(x = (x_1,x_2,x_3,x_4)\in\mathbb{R}^4\).
The density \(\rho(t,x)\) satisfies the same chemotactic PDE,
\[
\partial_t \rho
= \mu \Delta \rho
+ \chi \nabla_x \cdot\left(\rho \nabla_x (K * \rho)\right),
\]
with identical regularization and parameters as in the 3D laminar experiment in Section~4.
The background velocity is now a 3D laminar profile embedded in four dimensions:
\begin{equation}
    v(x;\sigma)
    = \sigma\bigl(e^{-(x_2^2 + x_3^2 + x_4^2)},\,0,\,0,\,0\bigr)^\top,
    \label{eq:4d_laminar}
\end{equation}
so that the advection acts purely along \(x_1\) while the remaining coordinates provide
cross--stream structure.  The interacting--particle approximation of the PDE becomes
\begin{equation}
    dX_j
    = -\frac{\chi}{J}\sum_{i\neq j}\nabla K_\delta(\lVert X_i - X_j\rVert)\,dt
      + v(X_j;\sigma)\,dt
      + \sqrt{2\mu}\,dW_j,
    \label{eq:4d_particle_dynamics}
\end{equation}
which is the natural 4D analogue of the 3D scheme in Section~4.
We implement~\eqref{eq:4d_particle_dynamics} exactly as in the 3D experiment, 
with the only modification that initialization is uniform on the 4D unit ball.

\paragraph{Evaluation metric in 4D}
Directly applying the histogram--based discrete \(W_2\) used in 3D
requires a \(24^4\) grid (331,776 cells), and the resulting cost matrix
would contain over \(10^{11}\) entries, which exceeds our 480\,GB RAM budget
even in single precision.  
Therefore, for each \(\sigma\) we estimate \(W_2\) using
\emph{two independent subsamples} of size \(m=2000\) drawn from the
reference and generated ensembles.  
We compute the Earth--Mover's Distance (EMD or 1-Wasserstein Distance) between these point clouds
with a squared Euclidean cost.  
This modification affects only the evaluation metric; 
both Meanflow and DP training remain unchanged and identical
to the 3D setting.

\begin{figure}[H]
\centering
\begin{minipage}{0.35\linewidth}
\centering
\scriptsize
\begin{tabular}{rcc}
\toprule
$\sigma$ & \textbf{Meanflow}  & \textbf{DP refinement} \\
\midrule
$20^{(\blacktriangle)}$  & 0.0038 & \textbf{0.0030} \\
$40^{(\blacktriangle)}$  & 0.0040 & \textbf{0.0031} \\
$60^{(\blacktriangle)}$ & 0.0040 & \textbf{0.0031} \\
$80^{(\blacktriangle)}$  & 0.0039 & \textbf{0.0032} \\
$100^{(\blacktriangle)}$ & 0.0044 & \textbf{0.0039} \\
$120^{(\blacktriangle)}$ & 0.0572 & \textbf{0.0036} \\
$140^{(\blacktriangle)}$ & 0.1011 & \textbf{0.0038} \\
$160^{(\bullet)}$ & 0.1129 & \textbf{0.0043} \\
$180^{(\bullet)}$ & 0.1431 & \textbf{0.0066} \\
$200^{(\bullet)}$ & 0.1526 & \textbf{0.0134} \\
\bottomrule
\end{tabular}
\end{minipage}
\hfill
\begin{minipage}{0.54\linewidth}
\centering
\includegraphics[width=0.95\linewidth]{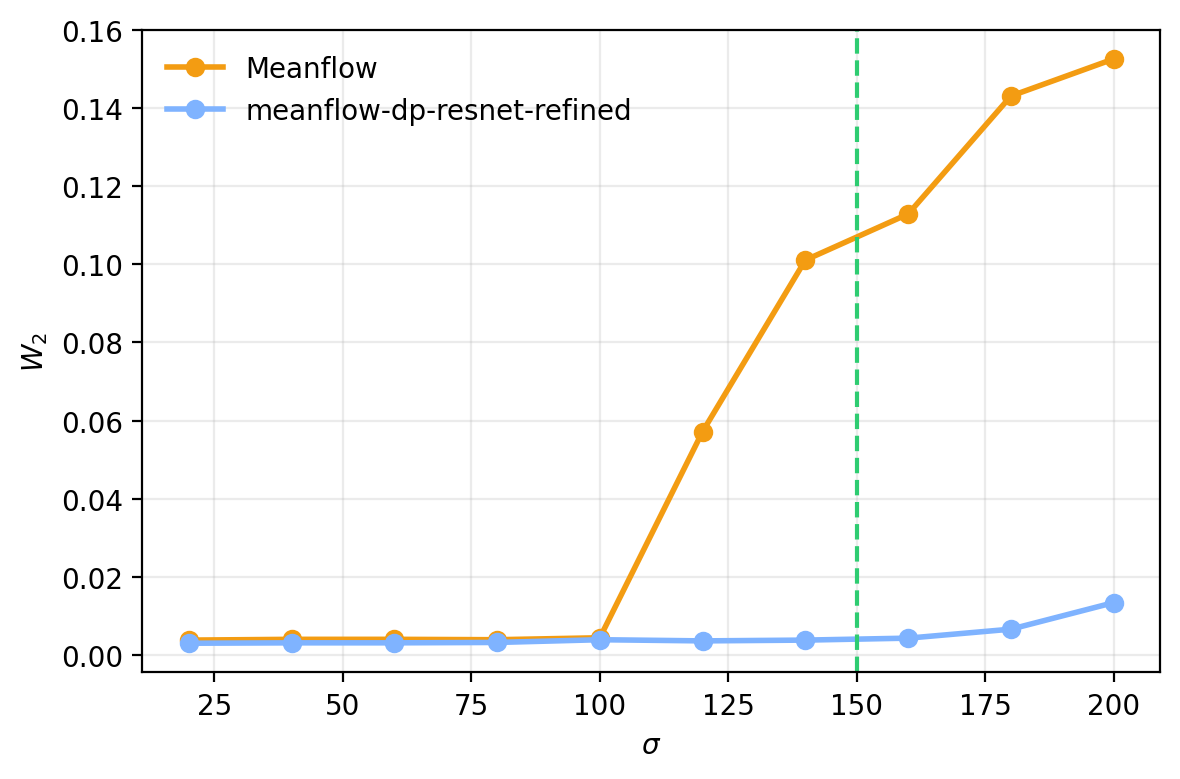}
\end{minipage}
\caption{4D Keller--Segel with laminar flow. 
Left: approximate \(W_2\) values for Meanflow and DP refinement across \(\sigma\).
Right: \(W_2\) vs.\ \(\sigma\) curves, showing Meanflow's error rapidly increasing
in the stiff advection regime while DP remains stable.}
\label{fig:4d_ks_4d}
\end{figure}

\paragraph{Results}
The left panel of Fig.~\ref{fig:4d_ks_4d} summarizes the approximate \(W_2\) values for
Meanflow and our DP refinement across \(\sigma\in\{20,\dots,200\}\).
In the moderate--advection range (\(\sigma\le 100\)),
both methods achieve errors around \(4\times 10^{-3}\), with DP consistently
slightly better.  As the system enters the singular--perturbation regime,
Meanflow's error grows sharply while DP remains nearly flat:
at \(\sigma=120\), Meanflow jumps to \(0.0572\) whereas DP stays at \(0.0036\)
(about \(16\times\) smaller); at \(\sigma=160\)--\(180\), the gap grows to 
\(20\times\)–\(26\times\); even at \(\sigma=200\), Meanflow reaches \(0.1526\)
while DP remains \(0.0134\).
The right panel of Fig.~\ref{fig:4d_ks_4d} illustrates the blow--up of Meanflow
beyond \(\sigma\approx 120\), contrasted with the stability of DP.\\
These results show that \emph{the advantage of our two--stage design becomes 
even more pronounced in higher dimensions}.  
\textbf{Stage~I}  provides a fast global displacement that moves particles
near the correct 4D support; \textbf{Stage~II} then performs a lightweight,
geometry--aware OT refinement that remains tractable on small subsamples 
yet robustly prevents the high--dimensional misalignment suffered by the
one--step generator.  
This behavior is consistent with the intuition established in the 3D analysis 
and further validates our claim that the ``coarse--to--fine'' Meanflow+DP
strategy is well suited for challenging high--dimensional transport problems.
 \subsection{Ablation Study: Batch Size and Coupling Refresh Rate}
\label{sec:ablation}
\noindent To assess the efficiency–accuracy trade-offs in Stage-II Wasserstein refinement, we perform an
ablation study over two key hyperparameters: the DP mini-batch size $B_{\mathrm{dp}}$ and the
coupling refresh period $\gamma_{\mathrm{renew}}$.  
Table~\ref{tab:ablation-batch-gamma} reports the resulting approximate
Wasserstein-2 distances across $\sigma \in \{20,\dots,200\}$, along with wall-clock time and peak
memory usage.
\paragraph{Effect of DP mini-batch size}
With $\gamma_{\mathrm{renew}}=50$ fixed, varying $B_{\mathrm{dp}}$ from $1500$ to $3500$ results in only
minor fluctuations in final $W_2$ values, where differences are typically on the order of $10^{-3}$ and
no configuration dominates uniformly.  
However, the computational cost scales nearly linearly with batch size, with total runtime increasing
from $893$~s at $1500$ to more than $3800$~s at $3500$, while peak memory remains unchanged.
Thus, a small batch size significantly accelerates training without sacrificing performance or
introducing memory overhead.  
We therefore adopt $B_{\mathrm{dp}}=1500$ (bold column in Table~\ref{tab:ablation-batch-gamma}) as our
default configuration.

\paragraph{Effect of coupling refresh rate}
Fixing the batch size to $1500$, we vary $\gamma_{\mathrm{renew}} \in \{5,25,50,100\}$.
Very frequent refreshes ($\gamma_{\mathrm{renew}}=5$) provide only negligible $W_2$ gains while being
$\sim 4\times$ slower, whereas overly sparse refreshes ($\gamma_{\mathrm{renew}}=100$) degrade accuracy,
especially at larger $\sigma$.  
The intermediate value $\gamma_{\mathrm{renew}}=50$ achieves the best balance between stability and
efficiency, reducing runtime to $854$~s while attaining nearly the lowest $W_2$ across the sweep,
with identical memory usage to all other settings.  
We therefore use $\gamma_{\mathrm{renew}}=50$ (bold column in Table~\ref{tab:ablation-batch-gamma})
for all main experiments, as it offers a substantial reduction in runtime while preserving accuracy
and memory consumption.

\begin{table}[h]
\centering
\caption{DP $W_2$ under different batch sizes and coupling refresh rates. 
Bold columns indicate the configuration adopted in the main experiments.}
\label{tab:ablation-batch-gamma}
\begin{minipage}{0.5\linewidth}
    \centering
    \scriptsize Batch size sweep ($\gamma_{\mathrm{renew}} = 50$)\\[0.3em]
    \begin{tabular}{c|ccccc}
    \hline
    $\sigma$ & \textbf{1500} & 2000 & 2500 & 3000 & 3500 \\
    \hline
    20  & \textbf{0.0047} & 0.0039 & 0.0044 & 0.0039 & 0.0039 \\
    40  & \textbf{0.0046} & 0.0038 & 0.0039 & 0.0040 & 0.0043 \\
    60  & \textbf{0.0049} & 0.0045 & 0.0040 & 0.0043 & 0.0043 \\
    80  & \textbf{0.0056} & 0.0042 & 0.0043 & 0.0047 & 0.0042 \\
    100 & \textbf{0.0059} & 0.0054 & 0.0050 & 0.0048 & 0.0045 \\
    120 & \textbf{0.0065} & 0.0053 & 0.0051 & 0.0056 & 0.0053 \\
    140 & \textbf{0.0073} & 0.0053 & 0.0053 & 0.0058 & 0.0056 \\
    160 & \textbf{0.0082} & 0.0064 & 0.0060 & 0.0058 & 0.0059 \\
    180 & \textbf{0.0140} & 0.0100 & 0.0095 & 0.0100 & 0.0102 \\
    200 & \textbf{0.0214} & 0.0166 & 0.0171 & 0.0172 & 0.0190 \\
    \hline
    time (s)    & \textbf{893.2} & 1395.0 & 2089.1 & 2892.9 & 3830.4 \\
    \hline
    \end{tabular}
\end{minipage}
\hfill
\begin{minipage}{0.47\linewidth}
    \centering
    \scriptsize Coupling refresh sweep (batch size $=1500$)\\[0.3em]
    \begin{tabular}{cccc}
    \hline
    5 & 25 & \textbf{50} & 100 \\
    \hline
    0.0038 & 0.0039 & \textbf{0.0047} & 0.0081 \\
    0.0043 & 0.0045 & \textbf{0.0046} & 0.0073 \\
    0.0045 & 0.0040 & \textbf{0.0049} & 0.0090 \\
    0.0046 & 0.0045 & \textbf{0.0056} & 0.0105 \\
    0.0049 & 0.0050 & \textbf{0.0059} & 0.0115 \\
    0.0054 & 0.0055 & \textbf{0.0065} & 0.0125 \\
    0.0052 & 0.0058 & \textbf{0.0073} & 0.0111 \\
    0.0064 & 0.0064 & \textbf{0.0082} & 0.0133 \\
    0.0102 & 0.0094 & \textbf{0.0140} & 0.0169 \\
    0.0163 & 0.0160 & \textbf{0.0214} & 0.0202 \\
    \hline
    3300.4 & 1109.5 & \textbf{893.2} & 771.5 \\
    \hline
    \end{tabular}
\end{minipage}
\label{tab1}
\end{table}

\subsection{Comparison with other baselines}
\label{sec:comparison-baselines}

\paragraph{IMM (self distillation)~\cite{zhou2025inductive}}
We implement an inductive moment matching (IMM) style one-step baseline adapted to the KS setting and closely related to the self-distilled flow-map view and IMM framework~\cite{zhou2025inductive,boffi2025build}.  For each physical parameter $\sigma$ on the same 8-point logarithmic grid used by Meanflow, we generate paired endpoints $(x_0, x_1)$ by simulating the KS dynamics from the shared initial configuration $x_0$ to the terminal state $x_1$. The model is a compact MLP (width 64, depth 4) that shares both architecture and $\sigma$-conditioning with the Meanflow network and parameterizes a time-conditioned flow map
\begin{equation}
    f_\theta(x, t \to s, \log_{10}\sigma)
  = 
x + (s-t)\,g_\theta(x,t,s,\log_{10}\sigma),
\end{equation}
so that a single forward pass predicts the displacement from any source time $t$ to any earlier target time $s$.
During training we sample a mini-batch of KS pairs for a randomly chosen $\sigma$ and draw nested times $0 \le s < r < t \le 1$.  We define a linear path between data and prior,
$x_t = (1-t)x_1 + t x_0$ and $x_r = (1-r)x_1 + r x_0$, and enforce temporal consistency by matching the distributions of the ``big-step'' map $f_\theta(x_t,t \to s)$ and the ``small-step'' map $f_\theta(x_r,r \to s)$.  These two sets of samples are partitioned into groups and compared using a Laplacian-kernel MMD loss whose bandwidth scales with the step size $|t-s|$, yielding the IMM-style objective
\begin{equation}
    \mathcal{L}_{\mathrm{IMM}}(\theta)
=
\mathbb{E}_{\sigma,(x_0,x_1),s<r<t}
\Big[
\mathrm{MMD}^2 \big(
f_\theta(x_t,t \to s),
f_\theta(x_r,r \to s)
\big)
\Big].
\end{equation}
Gradients flow only through the big-step branch while the midpoint branch is treated as a frozen teacher, implementing a self-distillation scheme without an external diffusion teacher.  All hyperparameters are kept identical to the Meanflow stage to ensure a fair comparison.

\begin{figure}[H]
    \centering
    \includegraphics[width=\linewidth]{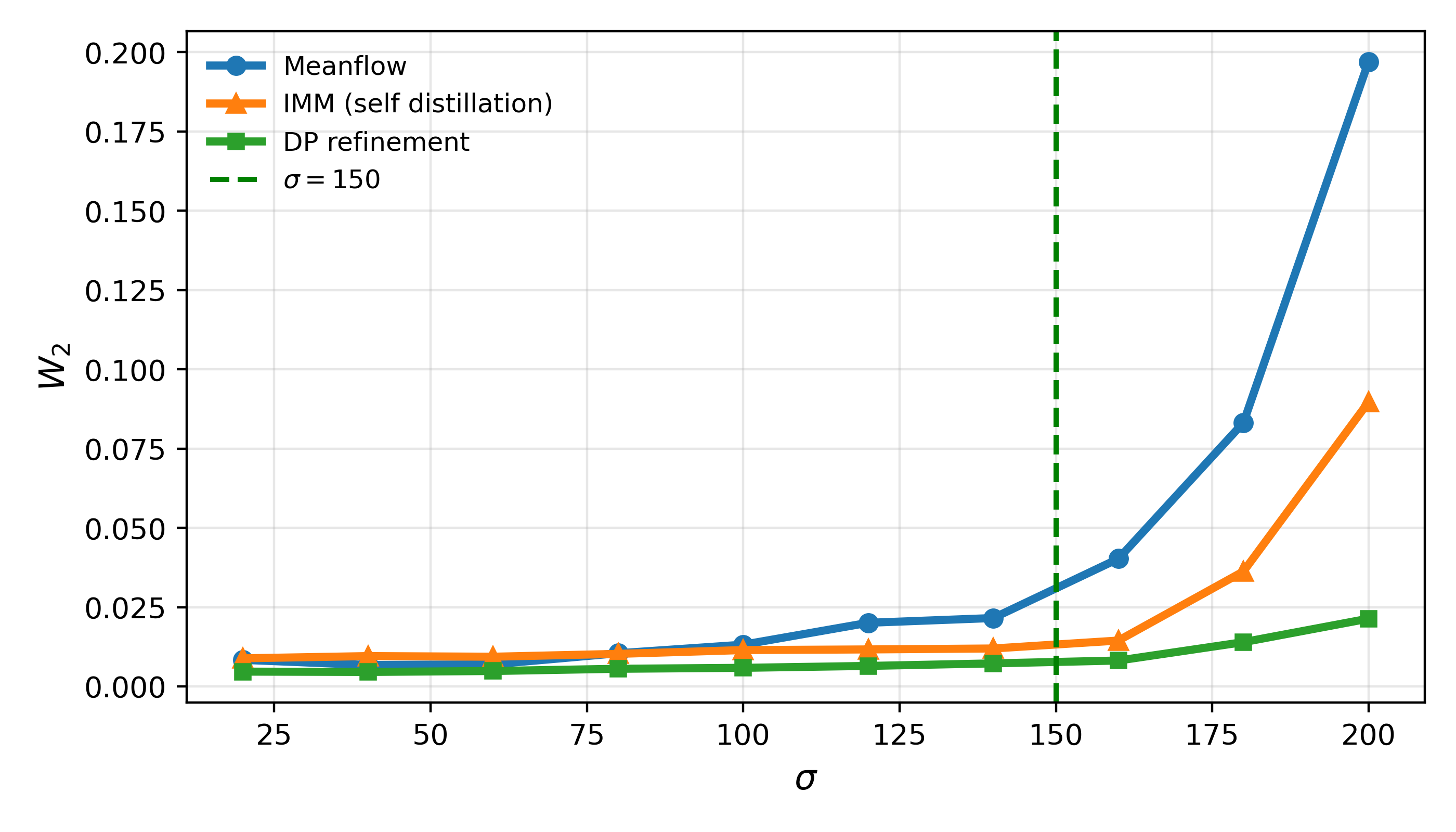}
    \caption{Comparison of the  2-Wasserstein metric $W_2$ across $\sigma$ for Meanflow, DP refinement, IMM (self distillation).}
    \label{fig:ks3d-baselines}
\end{figure}

\begin{figure}[H]
    \centering
    \caption{Qualitative comparison of 2D histograms at $\sigma=20$ (interpolation regime).  
    From top to bottom we show Meanflow, DP refinement, IMM (self distillation), and the KS reference solution.  
    All methods produce roughly isotropic particle clouds, but Meanflow exhibits slightly blurred density and a noticeable mismatch in the outer mass.  
    IMM reduce these distortions, while DP refinement yields the sharpest and most symmetric match to the reference across all coordinate projections.}
    \label{fig:ks3d-hist-20}
    \includegraphics[width=\linewidth]{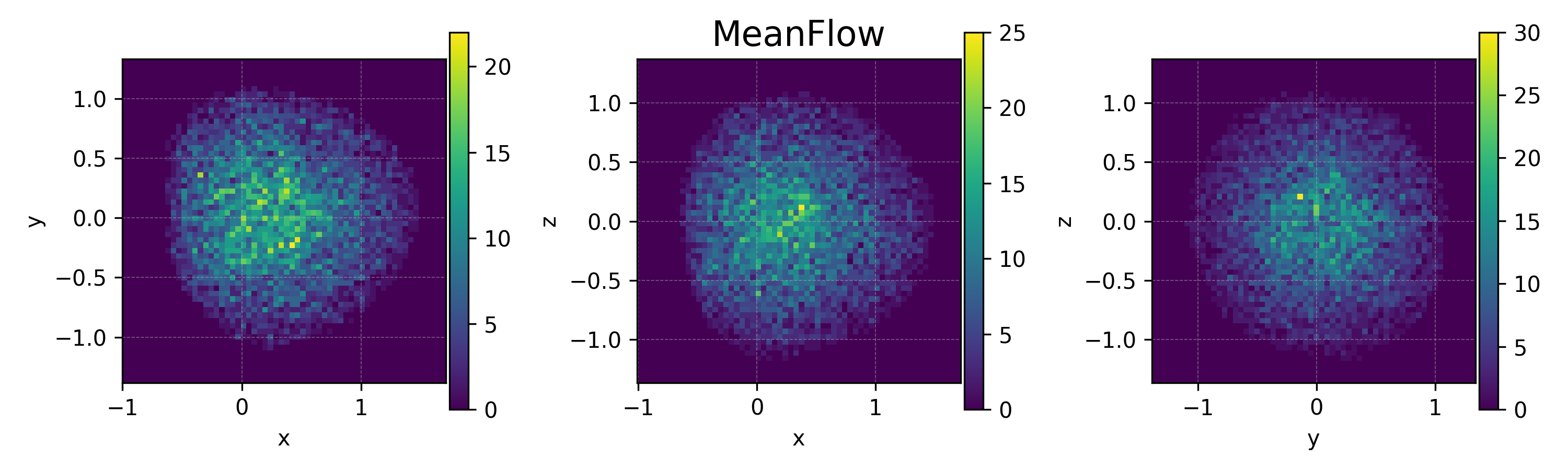}
    \includegraphics[width=\linewidth]{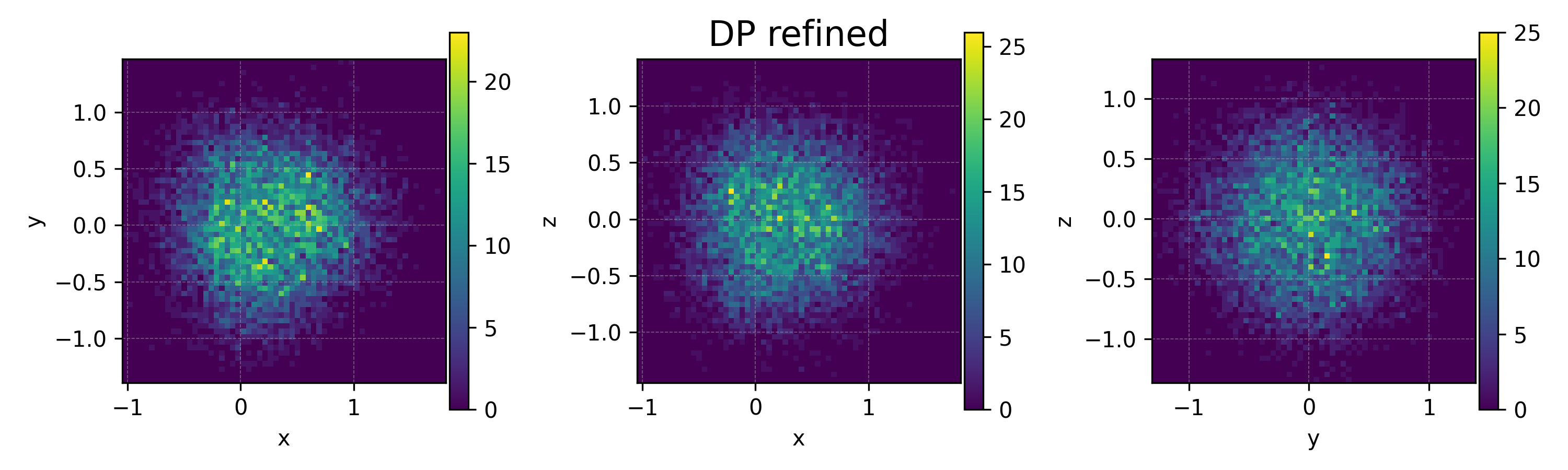}
    \includegraphics[width=\linewidth]{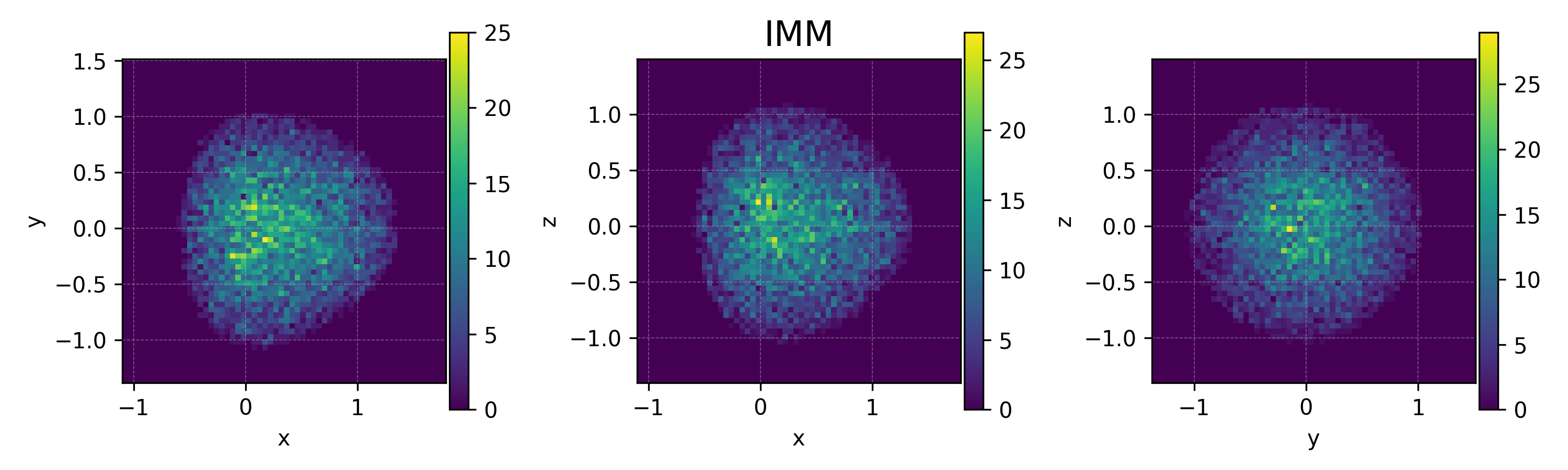}
    \includegraphics[width=\linewidth]{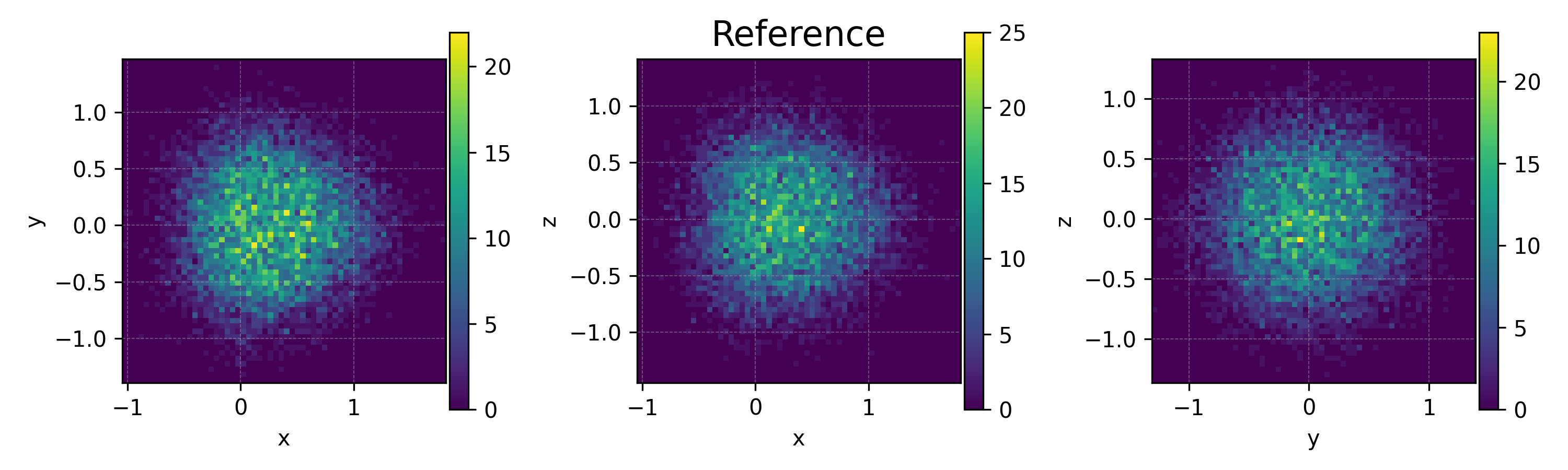}
\end{figure}

\begin{figure}[H]
    \centering
    \caption{Qualitative comparison of 2D histograms at $\sigma=160$ (extrapolation regime).  
    Meanflow significantly distorts the particle distribution, with mass shifted away from the reference and visible anisotropy.  
    IMM partially correct these errors but still exhibits broader, less concentrated densities.  
    In contrast, DP refinement closely tracks the reference histograms in all views, preserving both the core concentration and the shape of the support, confirming that it generalizes best beyond the training range.}
    \label{fig:ks3d-hist-160}
    \includegraphics[width=\linewidth]{meanflow_160.png}
    \includegraphics[width=\linewidth]{dp_160.png}
    \includegraphics[width=\linewidth]{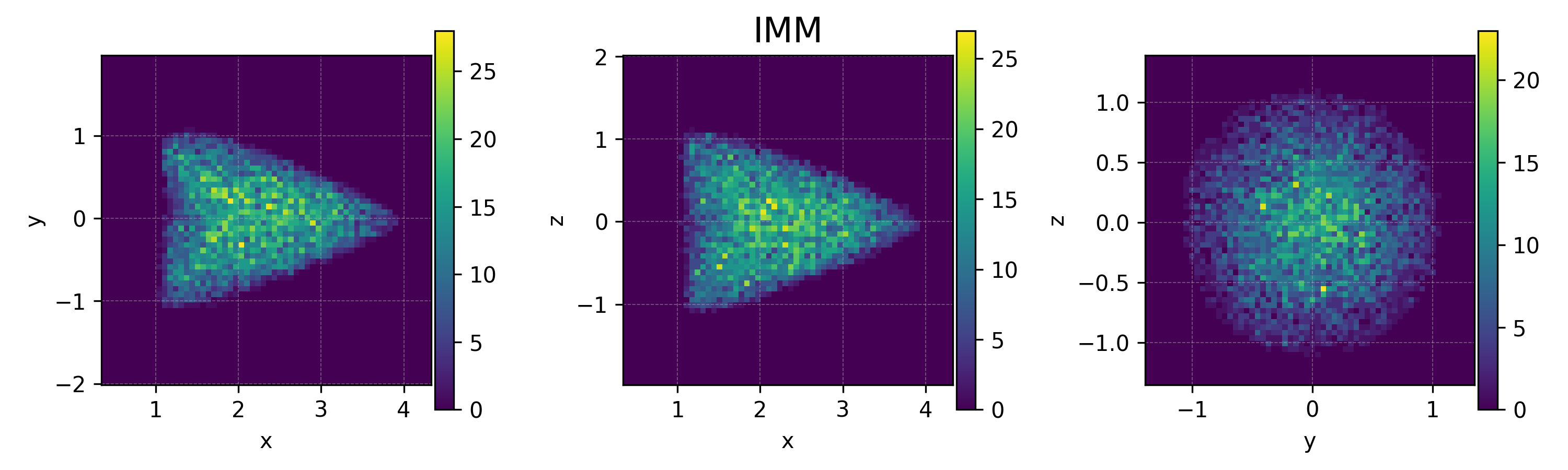}
    \includegraphics[width=\linewidth]{ref_160.png}
\end{figure}

\paragraph{Experiment analysis.}
Table~\ref{tab:ks3d-baselines} and Figure~\ref{fig:ks3d-baselines} compare Meanflow, IMM (self distillation), and our Two-step method (DP refinement) across $\sigma\in\{20,\dots,200\}$ using the approximate 2-Wasserstein metric $W_2$. 
Across the entire range, DP refinement achieves the lowest error, demonstrating that an additional refinement stage on top of a one-step proposal provides a consistent and effective correction.
In the interpolation/moderate regime ($\sigma\le 100$), all methods are reasonably accurate, but DP refinement already delivers a clear margin (e.g., $W_2=0.0047$ at $\sigma=20$ versus $0.0084/0.0089$ for Meanflow/IMM, and $0.0059$ at $\sigma=100$ versus $0.0132/0.0115$). 
As $\sigma$ increases into the more challenging regime (notably beyond the $\sigma=150$ marker in Figure~\ref{fig:ks3d-baselines}), the gap widens substantially: Meanflow’s error grows rapidly and IMM begins to drift, while DP refinement remains stable and significantly lower (e.g., $0.0082$ at $\sigma=160$ and $0.0214$ at $\sigma=200$, compared to $0.0145/0.0897$ for IMM and $0.0403/0.1970$ for Meanflow). 
Figures~\ref{fig:ks3d-hist-20} and~\ref{fig:ks3d-hist-160} corroborate these quantitative trends: at $\sigma=20$, DP refinement yields the sharpest and most symmetric match to the KS reference across projections, while at $\sigma=160$ it largely eliminates the mass shift and anisotropy visible in one-step baselines, preserving both the core concentration and the shape of the support. 
Overall, these results highlight the advantage of our two-step design: a refinement step that markedly improves accuracy and, crucially, strengthens extrapolation robustness.

\begin{table}[H]
\centering
\small
\caption{Approximate $W_2$ for different baselines on the 3D KS--Laminar experiment.}
\label{tab:ks3d-baselines}
\begin{tabular}{r|ccc}
\toprule
$\sigma$ & Meanflow & IMM (self distillation) & \textbf{DP refinement} \\
\midrule
20  & 0.0084 & 0.0089 & \textbf{0.0047} \\
40  & 0.0068 & 0.0096 & \textbf{0.0046} \\
60  & 0.0070 & 0.0094 & \textbf{0.0049} \\
80  & 0.0105 & 0.0103 & \textbf{0.0056} \\
100 & 0.0132 & 0.0115 & \textbf{0.0059} \\
120 & 0.0201 & 0.0117 & \textbf{0.0065} \\
140 & 0.0216 & 0.0120 & \textbf{0.0073} \\
160 & 0.0403 & 0.0145 & \textbf{0.0082} \\
180 & 0.0832 & 0.0363 & \textbf{0.0140} \\
200 & 0.1970 & 0.0897 & \textbf{0.0214} \\
\bottomrule
\end{tabular}
\end{table}


\subsection{Comparison with Regularized Sinkhorn}
\label{sec:sinkhorn}
\begin{figure}[H]
    \centering
    \caption{Qualitative comparison of Mini-batch OT refinement under different Sinkhorn regularization coefficients versus exact EMD refinement when $\sigma=160$. Large coefficients oversmooth the distribution; small coefficients improve accuracy but become extremely slow. EMD achieves the cleanest alignment at the lowest computational cost.}
    \label{fig:sinkhorn_vs_emd_qualitative}
    \includegraphics[width=\linewidth]{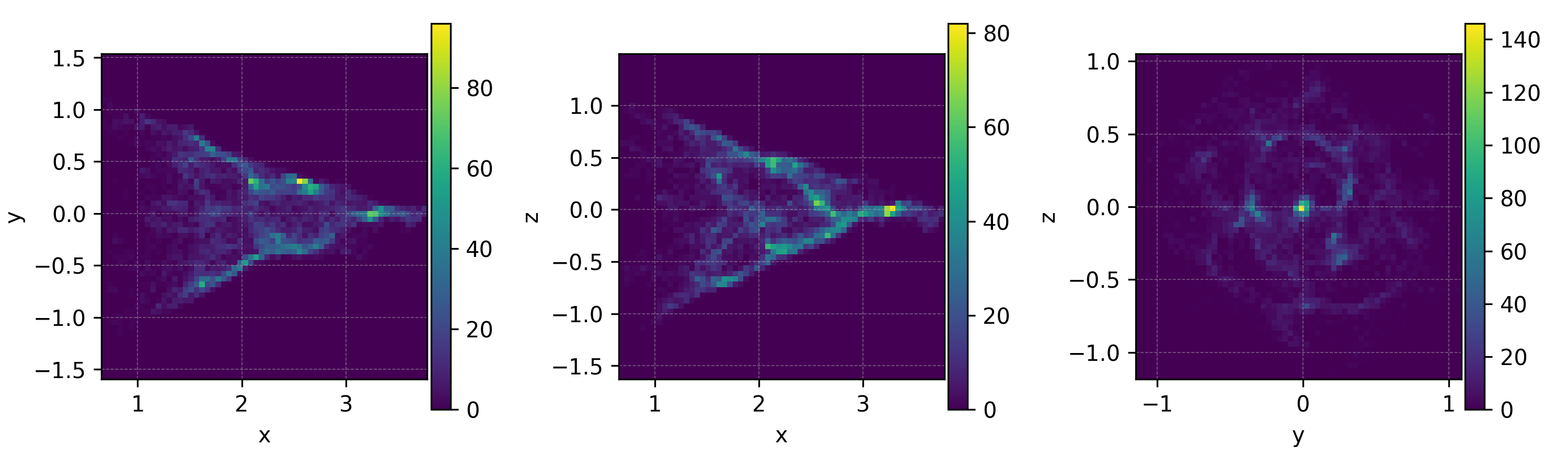}
        \vspace{-2.5em}
    \subcaption{$\varepsilon=0.05$}
     \vspace{-0.7em}
    \includegraphics[width=\linewidth]{sink5e-2.png}
        \vspace{-2.5em}
    \subcaption{$\varepsilon=0.01$}
     \vspace{-0.7em}
    \includegraphics[width=\linewidth]{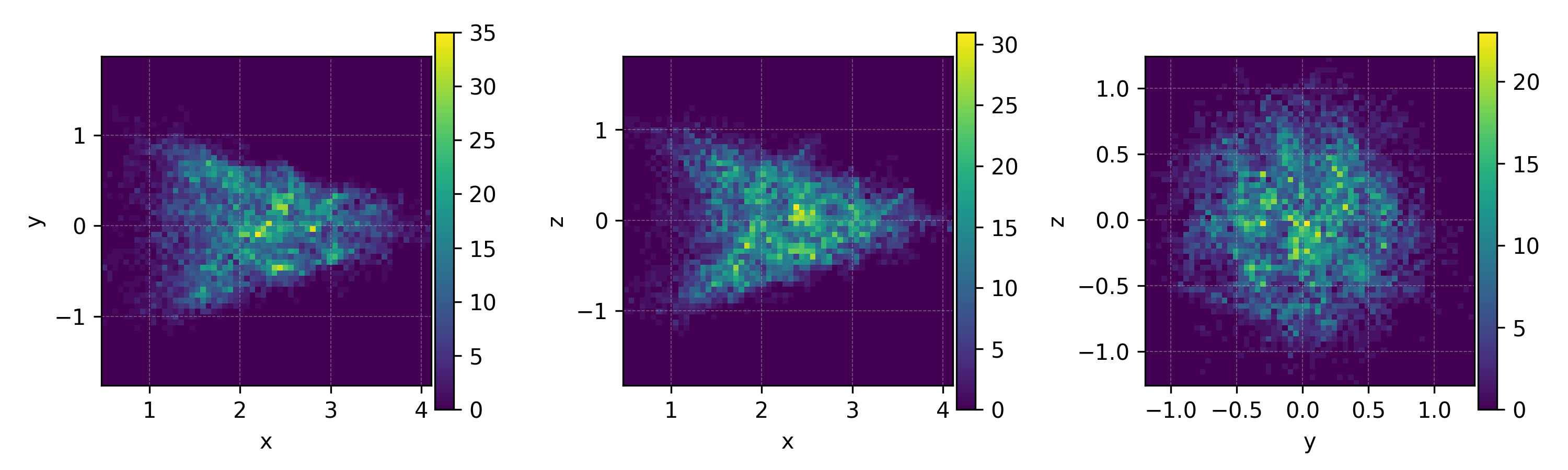}
        \ \vspace{-2.5em}
    \subcaption{$\varepsilon=0.005$}
     \vspace{-0.7em}
    \includegraphics[width=\linewidth]{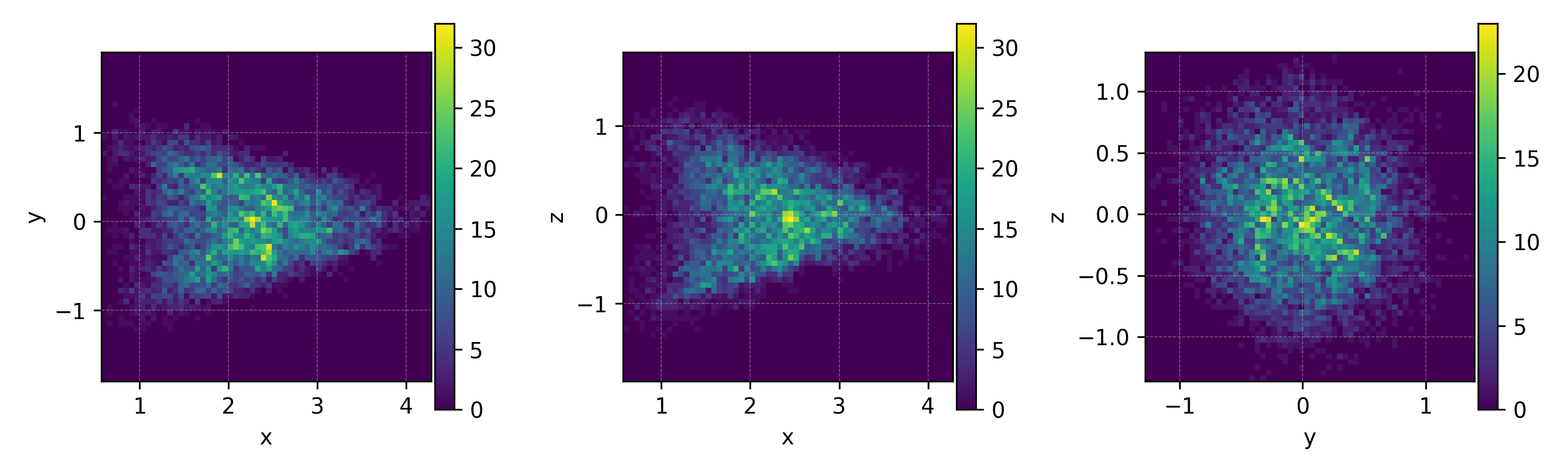}
        \vspace{-2.5em}
    \subcaption{$\varepsilon=0.001$}
     \vspace{-0.7em}
    \includegraphics[width=\linewidth]{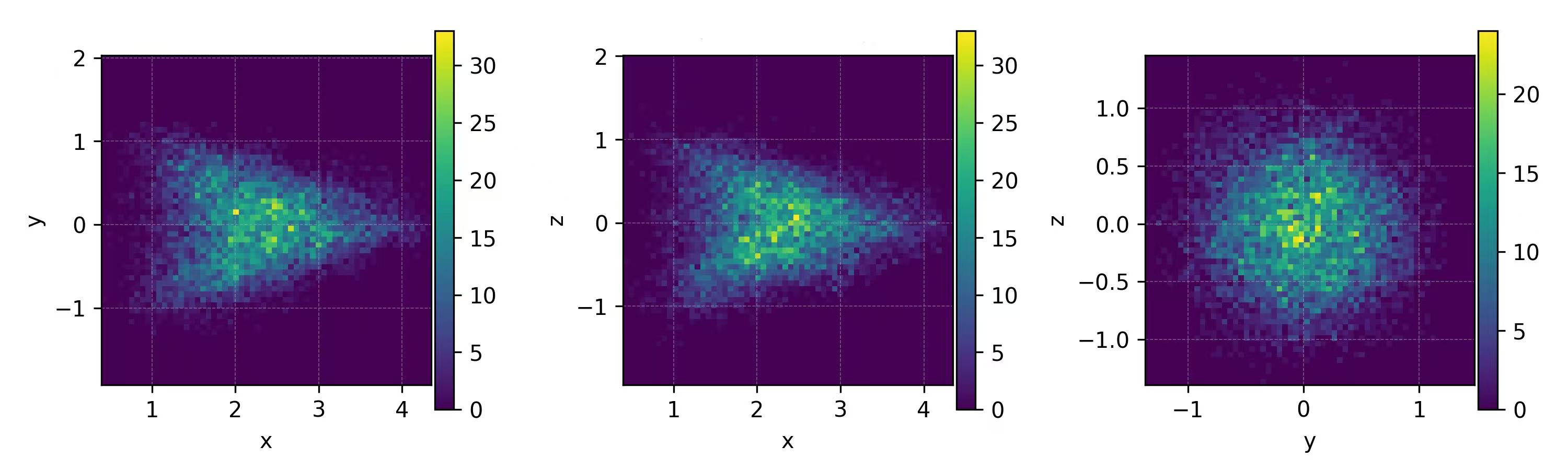}
         \vspace{-2.3em}
    \subcaption{EMD solver}
\end{figure}

\noindent We compare our exact mini-batch EMD refinement with entropically
regularized Sinkhorn solvers using coefficients $\varepsilon \in \{0.05, 0.01, 0.005, 0.001\}$.
We keep all components of the two-stage pipeline fixed and only change the OT solver in \textbf{Stage~II}. Compared with Algorithm~2, we only replace the coupling-update step.
Instead of solving the exact mini-batch EMD problem
\begin{equation}
\gamma^{\star}
=
\arg\min_{\gamma \in \Gamma_{B}}
\langle C, \gamma \rangle ,
\label{eq:emd_coupling}
\end{equation}
we compute an entropically regularized Sinkhorn coupling
\begin{equation}
\gamma_{\varepsilon}
=
\arg\min_{\gamma \in \Gamma_{B}}
\left(
\langle C, \gamma \rangle
+
\varepsilon \sum_{i,j}
\gamma_{ij}\bigl(\log \gamma_{ij} - 1\bigr)
\right),
\label{eq:sinkhorn_coupling}
\end{equation}
where $\varepsilon > 0$ is the entropic regularization coefficient. All other steps in Algorithm~2 remain unchanged.\\
Table~\ref{tab:sinkhorn_vs_emd} reports the resulting approximate $W_2$
across $\sigma \in \{20,\dots,200\}$ along with runtime and memory usage, while 
Figure~\ref{fig:sinkhorn_vs_emd_qualitative} provides qualitative comparisons of the 
learned particle distributions under each regularization strength. Across all coefficients, Sinkhorn achieves reasonable alignment but exhibits a clear
accuracy--efficiency trade-off. Larger entropic regularization ($\varepsilon=0.05$ and
$\varepsilon=0.01$) leads to excessive smoothing of the transport plan, producing visibly
blurred distributions and noticeably higher $W_2$ values, especially in the
singular-perturbation regime ($\sigma \ge 150$). Reducing the coefficient improves
accuracy, and $\varepsilon = 0.001$ performs closest to EMD. However, the computational
cost increases dramatically as $\varepsilon$ becomes small: the total runtime grows from 
$6538$~s at $\varepsilon=0.05$ to more than $6.2\times 10^4$~s at $\varepsilon=0.001$,
while memory usage remains higher than that of EMD.\\
In contrast, EMD refinement simultaneously achieves:
(i) the lowest or near-lowest $W_2$ across all $\sigma$;
(ii) an order-of-magnitude speedup compared to even the fastest Sinkhorn run; 
(iii) the smallest memory footprint.
These results indicate that, after Meanflow has concentrated mass on the correct support,
the OT geometry becomes sufficiently well-conditioned that exact EMD is both more accurate
and dramatically more efficient than Sinkhorn. We therefore adopt EMD as the default
mini-batch OT solver in all main experiments.

\begin{table}[H]
    \centering
    \caption{Comparison of $W_2$ under different regularized Sinkhorn coefficients. 
    EMD refinement (END-RF) is both significantly faster and yields the best overall accuracy.}
    \label{tab:sinkhorn_vs_emd}
    \begin{tabular}{c|ccccc}
        \hline
        $\sigma$ & 0.001 & 0.005 & 0.01 & 0.05 & EMD-RF \\
        \hline
        20  & 0.003961 & 0.004436 & 0.004859 & 0.007994 & 0.0047 \\
        40  & 0.004126 & 0.004357 & 0.004747 & 0.007871 & 0.0046 \\
        60  & 0.003915 & 0.004419 & 0.004765 & 0.008195 & 0.0049 \\
        80  & 0.004640 & 0.005143 & 0.005538 & 0.009247 & 0.0056 \\
        100 & 0.004764 & 0.004913 & 0.006174 & 0.009604 & 0.0059 \\
        120 & 0.005479 & 0.005843 & 0.007758 & 0.010074 & 0.0065 \\
        140 & 0.005214 & 0.006203 & 0.008551 & 0.010624 & 0.0073 \\
        160 & 0.006398 & 0.006902 & 0.009029 & 0.011593 & 0.0082 \\
        180 & 0.010744 & 0.012127 & 0.020977 & 0.025443 & 0.0140 \\
        200 & 0.015145 & 0.017245 & 0.027149 & 0.039835 & 0.0214 \\
        \hline
        time (s)    & 62879.88 & 44889.60 & 26305.99 & 6537.78 & 893.2 \\
        memory (MB) & 1479.76  & 1445.29  & 1504.32  & 1442.92 & 1154.7 \\
        \hline
    \end{tabular}
\end{table}

\subsection{Comparison with multi-step Meanflow Results}
\label{sec:app:multistep-meanflow}
\begin{figure}[H]
    \centering
    \includegraphics[width=1\linewidth]{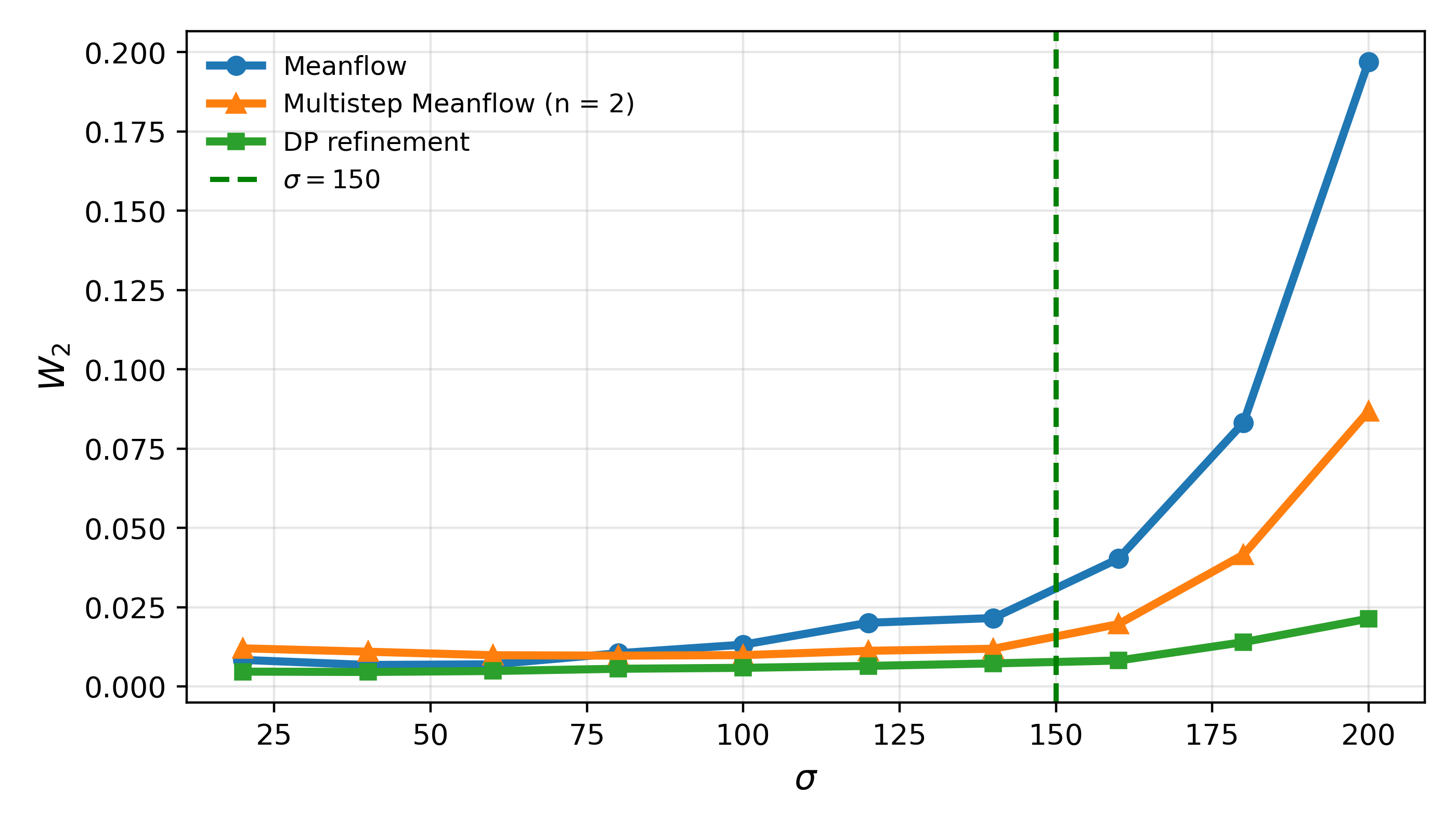}
    \caption{Empirical $W_2$ versus advection strength $\sigma$ for one-step Meanflow, multi-step Meanflow ($n=2$), and DP refinement. The dashed vertical line marks the end of the training range ($\sigma=150$). DP refinement remains low and stable in extrapolation and outperforms multi-step Meanflow across the full sweep.}
  \label{fig:ks-laminar-multistep-meanflow-w2}
\end{figure}

\noindent To test whether additional Meanflow steps alone can close the accuracy gap, we also evaluate a simple \emph{multi-step Meanflow} baseline that composes the Meanflow transport for $n=2$ steps. Table~\ref{tab:ks-laminar-multistep-meanflow} and Fig.~\ref{fig:ks-laminar-multistep-meanflow-w2} report the empirical $W_2$ on the 3D KS--Laminar benchmark.
While using two Meanflow steps can mildly reduce error in part of the moderate regime, the improvement is not consistent and can even degrade performance due to error accumulation (e.g., $\sigma=20$: $0.0084 \rightarrow 0.0120$, and $\sigma=160$: $0.0403 \rightarrow 0.0416$). In contrast, our Two-Step Diffusion (Meanflow + DP refinement) consistently achieves the lowest $W_2$ across all $\sigma$ and is substantially more robust in extrapolation. For example, at $\sigma=160/180/200$, multi-step Meanflow attains $W_2=0.0416/0.0565/0.0868$, whereas DP refinement reduces these to $0.0082/0.0140/0.0214$ (about $5.1\times/4.0\times/4.1\times$ smaller, respectively). This indicates that simply increasing Meanflow steps does not reliably correct distributional misalignment, whereas the near-identity DP corrector provides a targeted refinement that improves mass placement and stabilizes performance beyond the training range.
\begin{table}[h]
    \centering
    \begin{tabular}{lccc}
\toprule
\textbf{$\sigma$} 
& \textbf{Meanflow} 
& \textbf{Multi-step Meanflow (n = 2)} 
& \textbf{DP refinement} \\
\midrule
$20^{(\blacktriangle)}$  & 0.0084 & 0.0120 & \textbf{0.0047} \\
$40^{(\blacktriangle)}$  & 0.0068 & 0.0110 & \textbf{0.0046} \\
$60^{(\blacktriangle)}$  & 0.0070 & 0.0099 & \textbf{0.0049} \\
$80^{(\blacktriangle)}$  & 0.0105 & 0.0097 & \textbf{0.0056} \\
$100^{(\blacktriangle)}$ & 0.0132 & 0.0099 & \textbf{0.0059} \\
$120^{(\blacktriangle)}$ & 0.0201 & 0.0113 & \textbf{0.0065} \\
$140^{(\blacktriangle)}$ & 0.0216 & 0.0119 & \textbf{0.0073} \\
$160^{(\bullet)}$        & 0.0403 & 0.0416 & \textbf{0.0082} \\
$180^{(\bullet)}$        & 0.0832 & 0.0565 & \textbf{0.0140} \\
$200^{(\bullet)}$        & 0.1970 & 0.0868 & \textbf{0.0214} \\
\bottomrule

\end{tabular}
    \caption{Empirical $W_2$ on the 3D KS--Laminar benchmark comparing one-step Meanflow, multi-step Meanflow ($n=2$), and our DP refinement (Two-Step Diffusion). $\blacktriangle$ denotes interpolation and $\bullet$ denotes extrapolation.}
  \label{tab:ks-laminar-multistep-meanflow}
\end{table}

\begin{figure}[H]
 \centering
     \includegraphics[width=\linewidth]{meanflow_160.png}
     \includegraphics[width=\linewidth]{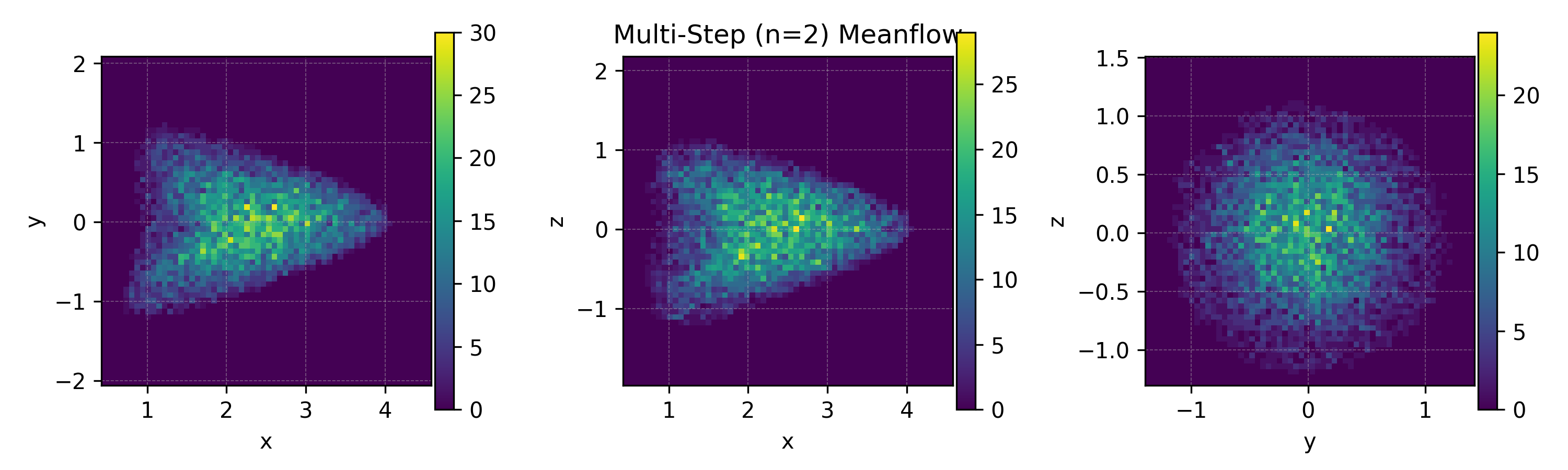}
     \includegraphics[width=\linewidth]{dp_160.png}
     \includegraphics[width=\linewidth]{ref_160.png}
       \caption{Qualitative comparison of 2D projections at $\sigma=160$ (extrapolation) for one-step Meanflow, multi-step Meanflow ($n=2$), DP refinement, and the KS reference. Multi-step Meanflow remains relatively diffusive with residual bias, while DP refinement better matches the reference support and anisotropy.}
  \label{fig:ks-laminar-multistep-meanflow-qual}
\end{figure}

\bibliographystyle{elsarticle-num}
\bibliography{cas-refs}

\end{document}